\documentclass[11pt]{article}
\usepackage{amsmath,amssymb}
\usepackage[all]{xy}
\input diagxy

\font\fr=eufm10 scaled \magstep 1 %(caracteres goticos)
\newtheorem{teor}{Theorem}
\newtheorem{prop}{Proposition}

\newtheorem{definition}{Definition}
\newtheorem{state}{Statement}

\def\beq{\begin{equation}}
\def\eeq{\end{equation}}
\def\bea{\begin{eqnarray}}
\def\eea{\end{eqnarray}}
\def\beann{\begin{eqnarray*}}
\def\eeann{\end{eqnarray*}}
\def\beasn{\begin{sneqnarray}}
\def\eeasn{\end{sneqnarray}}
\def\ben{\begin{enumerate}}
\def\een{\end{enumerate}}
\def\bit{\begin{itemize}}
\def\eit{\end{itemize}}
\def\proof{ (\emph{Proof\/}) }
\newcommand{\ds}{\displaystyle}
\def\derpar#1#2{\frac{\partial{#1}}{\partial{#2}}}

\def\moment#1#2#3{{#1}_{#2}, \ldots, {#1}_{#3}}
\def\qed{\ifvmode\removelastskip\fi
{\unskip\nobreak\hfil\penalty50\hbox{}\nobreak\hfil
\hbox{\vrule height1.2ex width1.2ex}\parfillskip=0pt
\finalhyphendemerits=0 \par\smallskip}}
\def\vf{\mbox{\fr X}}
\def\df{{\mit\Omega}}
\def\Lag{{\bf L}}

\def\Lden{{\cal L}}
\def\costf{\mathbb{L}}

\def\d{{\rm d}}

\def\Real{\mathbb{R}}
\def\R{\mathbb{R}}

\def\Tan{{\rm T}}

\def\inn{\mathop{i}\nolimits}
\def\Cinfty{{\rm C}^\infty}
\def\tabaddress#1{{\small\it\begin{tabular}[t]{c}#1
\\[1.2ex]\end{tabular}}}

\title{SKINNER-RUSK UNIFIED FORMALISM\\
  FOR OPTIMAL CONTROL SYSTEMS AND APPLICATIONS}
\author{\sc Mar\'\i a Barbero-Li\~n\'an\thanks{{\bf e}-{\it mail}:
  mbarbero@ma4.upc.edu},
  Arturo Echeverr\'\i a-Enr\'\i quez\thanks{{\bf e}-{\it mail}:
  arturo@ma4.upc.edu},
  \\
  \tabaddress{Departamento de Matem\'atica Aplicada IV\\
  Edificio C-3, Campus Norte UPC\\
  C/ Jordi Girona 1. 08034 Barcelona. Spain}
  \\
{\sc David Mart\'\i n de Diego\thanks{{\bf e}-{\it mail}:
d.martin@imaff.cfmac.csic.es}} \\
\tabaddress{Instituto de Matem\'aticas y F\'\i sica Fundamental, CSIC\\
   C/ Serrano 123. 28006 Madrid. Spain}
     \\
{\sc Miguel C. Mu\~noz-Lecanda\thanks{{\bf e}-{\it mail}:
  matmcml@ma4.upc.edu}},
{\sc Narciso Rom\'an-Roy\thanks{{\bf e}-{\it mail}:
  nrr@ma4.upc.edu}},
  \\
  \tabaddress{Departamento de Matem\'atica Aplicada IV\\
   Edificio C-3, Campus Norte UPC\\
   C/ Jordi Girona 1. 08034 Barcelona. Spain}}
\date{{\sl J. Phys. A: Math. Theor.} {\bf 40} (2007) 12071--12093}

\begin{document}

\maketitle

\pagestyle{myheadings}

\thispagestyle{empty}

\begin{abstract}
A geometric approach to time-dependent optimal control
problems is proposed. This formulation is based on  the
Skinner and Rusk formalism  for Lagrangian and Hamiltonian
systems. The corresponding unified formalism developed for
optimal control systems allows us to formulate
geometrically the necessary conditions given by a weak form
of Pontryagin's Maximum Principle, provided that the
differentiability with respect to controls is assumed and
the space of controls is open. Furthermore, our method is
also valid for implicit optimal control systems and, in
particular, for the so-called descriptor systems (optimal
control problems including both differential and algebraic
equations).
\end{abstract}


\bigskip

\vbox{\raggedleft AMS s.\,c.\,(2000): 70G45, 49J15, 34A26, 49K15, 70H03, 70H05 }\null

\markright{\sc M. Barbero-Li\~n\'an {\it et al\/},
    \sl Skinner-Rusk unified formalism...}

  \clearpage

\tableofcontents

\newpage

\section{Introduction}

In 1983 Skinner and Rusk introduced a representation
of the dynamics of an autonomous mechanical system which combines
the Lagrangian and Hamiltonian features \cite{SR-83}. Briefly,  in this
formulation, one  starts with a differentiable manifold $Q$ as the
configuration space, and the Whitney sum $TQ\oplus T^*Q$ as the
evolution space (with canonical projections $\rho_1: TQ \oplus
T^*Q\longrightarrow TQ$ and $\rho_2: TQ \oplus T^*Q\longrightarrow
T^*Q$).  Define on $TQ\oplus T^*Q$ the presymplectic 2-form
$\Omega=\rho_2^*\omega_Q$, where $\omega_Q$ is the canonical
symplectic form on $T^*Q$, and observe that the rank of this
presymplectic form is everywhere equal to $2n$. If the dynamical
system under consideration admits a Lagrangian description, with
Lagrangian $L\in C^{\infty}(TQ)$, then we obtain a
(presymplectic)-Hamiltonian representation on $TQ\oplus T^*Q$
given by the presymplectic 2-form $\Omega$ and the Hamiltonian
function $ H=\langle \rho_1,\rho_2\rangle -\rho_1^* L \; , $ where
$\langle \cdot , \cdot \rangle$ denotes the natural pairing
between vectors and covectors on $Q$.  In this Hamiltonian system
the dynamics is given by vector fields $X$, which are solutions to
the Hamiltonian equation $\inn(X)\Omega=dH$. If $L$ is regular, then
there exists a unique vector field $X$ solution to the previous
equation, which is tangent to the graph of the Legendre map
 ${\cal F}L\colon TQ\longrightarrow T^*Q$. In the singular case, it is necessary
to develop a constraint algorithm in order to find a submanifold
(if it exists) where there exists a well-defined dynamical vector field.

The idea of this formulation was to obtain a common
framework for both regular and singular dynamics,
obtaining simultaneously the Hamiltonian and Lagrangian formulations of the
dynamics. Over the years, however, Skinner and
Rusk's framework was extended in many directions. For instance, Cantrijn {\it
et al} \cite{CMC-2002} extended this formalism for explicit
time-dependent systems using a jet bundle language; Cort\'es {\it
et al} \cite{CLMM-2002} use the Skinner and Rusk formalism to
consider vakonomic mechanics and the comparison between the
solutions of vakonomic and nonholonomic mechanics. In
\cite{ELMMR-04,LMM-2002,RRS} the authors developed the Skinner-Rusk
model for classical field theories.

Furthermore,  the Skinner-Rusk formalism seems to be a natural geometric
setting for Pontryagin's Maximum Principle. In this paper,
whose roots are in the developments made in
\cite{CMC-2002,ELMMR-04,LMM-2002}, we adapt the Skinner-Rusk formalism to study
time-dependent optimal control problems.
In this way we obtain a geometric version of the Maximum Principle
that can be applied to a wide range of control systems.
For instance, these techniques enables to tackle geometrically
 implicit optimal control systems,  that is, those where the
control equations are implicit. In fact, systems of
differential-algebraic equations appear frequently in control theory.
 Usually, in the literature, it
is assumed that it is possible to rewrite the problem as an
explicit system of differential equations, perhaps using the
algebraic conditions to eliminate some variables, as in the case of
holonomic constraints. However, in general, a control system is
described as a system of equations of the type $F(t, x, \dot{x},
u)=0$, where the $x$'s denote the state variables and the $u$'s the
control variables, and there are some interesting cases where
the system is not described by the traditional equations
$\dot{x}=G(t, x, u)$.

The main results of this work can be found in Sections \ref{section-control} and \ref{iocp},
where we give a general method to deal with explicit and implicit systems.
As examples, we consider the case of optimal
control of Lagrangian mechanical systems (see \cite{AF,BOV-2002,B,BLM-2000}) and also
optimal control for descriptor systems \cite{muller,M-1999}.
Both examples have significant engineering applications.

The organization of the paper is as follows: Section
\ref{srfnas} is devoted to giving an alternative approach
of the Skinner-Rusk formalism for time dependent mechanical
systems. In Section 3 we develop the unified formalism for
explicit time-dependent optimal control problems giving a geometric Pontryagin's
Maximum Principle in a weak form, and in
Section 4 we do the same for implicit optimal control systems. Section 5
is devoted to examples and applications: first we study the
optimal control of Lagrangian systems with controls; that
is, systems defined by a Lagrangian and external forces
depending on controls \cite{AF,BOV-2002,B,BLM-2000}.
 These are considered as implicit systems defined by the
Euler-Lagrange equations. Second, we analyze a quadratic
optimal control problem for a descriptor system
\cite{muller}. We point out the importance of these kinds of
systems in engineering problems \cite{M-1999} and
references therein. Finally, we include an Appendix where
geometric features about Tulczyjew's operators, contact
systems and the Euler-Lagrange equations for forced
systems are explained.

All the manifolds are real, second countable and ${\cal
C}^{\infty}$. The maps are assumed to be ${\cal
C}^{\infty}$. Sum over repeated indices is understood.

\section{Skinner-Rusk unified formalism for non-autonomous systems}
\protect\label{srfnas}

This formalism is a particular case of the unified formalism for
field theories developed in \cite{ELMMR-04} and also in
\cite{LMM-2002}. See \cite{CMC-2002} for an alternative but
equivalent approach, and \cite{GM-05} for an extension of this
formalism to other kinds of more general time-dependent singular
differential equations.

%\subsection{Unified formalism}

In the jet bundle description of non-autonomous dynamical systems,
the configuration bundle is $\pi\colon E\to\Real$, where $E$ is a
$(n+1)$-dimensional differentiable manifold endowed with local
coordinates $(t,q^i)$, and $\Real$ has as a global coordinate $t$.
The jet bundle of local sections of $\pi$, $J^1\pi$, is the {\sl
velocity phase space} of the system, with natural coordinates
$(t,q^i,v^i)$, adapted to the bundle $\pi\colon E\to\Real$, and
natural projections
$$
\pi^1 \colon J^1\pi \to E \quad ,\quad \bar\pi^1 \colon J^1\pi \to
\Real \, .
$$

A Lagrangian density $\Lden\in\df^1(J^1\pi)$ is a
$\bar\pi^1$-semibasic $1$-form on $J^1\pi$, and it is usually
written as $\Lden = L\,\d t$, where $L\in\Cinfty (J^1\pi)$ is the
{\sl Lagrangian function} determined by $\Lden$. Throughout this paper
we denote by $\d t$ the volume form in $\Real$, and its pull-backs
to all the manifolds.

The canonical structure of the bundle $J^1\pi$
allows us to define the {\sl Poincar\'e-Cartan forms} associated with the
Lagrangian density $\Lden$,
and then the Euler-Lagrange equations are written intrinsically
(see \cite{EMR-91,Sa-89}).

Furthermore, we have the {\sl extended momentum phase space} $\Tan^*E$,
and the {\sl restricted momentum phase space} which is defined by
$J^1\pi^*= \Tan^*E/\pi^*\Tan^*\Real$. Local coordinates in these
manifolds are $(t,q^i,p,p_i)$ and $(t,q^i,p_i)$, respectively.
Then, the following natural projections are
$$
\tau^1 \colon J^1\pi^*\to E \quad ,\quad \bar\tau^1=\pi\circ\tau^1
\colon J^1\pi^* \to \Real \quad , \quad \mu \colon\Tan^*E \to
J^1\pi^* \quad , \quad p\colon\Tan^*E \to \Real \, .
$$
Let $\Theta\in\df^1(\Tan^*E)$ and
$\Omega=-\d\Theta\in\df^2(\Tan^*E)$ be
the canonical forms of $\Tan^*E$ whose local expressions are
$$
\Theta=p_i\d q^i+ p\d t \quad ,\quad \Omega=\d q^i\wedge\d p_i+\d
t\wedge\d p \, .
$$
The Hamilton equations can be written intrinsically
from these canonical structures
(see, for instance, \cite{EMR-91,Ku-tdms,MS-98,Ra1,St-2005}).

Now we introduce the geometric framework for the
unified Skinner-Rusk formalism for non-autonomous systems.
We define the {\sl extended jet-momentum bundle} ${\cal W}$ and
the {\sl restricted jet-momentum bundle} ${\cal W}_r$
$$
{\cal W}= J^1\pi \times_{E}\Tan^*E
\quad , \quad
{\cal W}_r =  J^1\pi \times_{E}  J^1\pi^*
$$
with natural coordinates $(t,q^i,v^i,p,p_i)$ and $(t,q^i,v^i,p_i)$,
respectively. We have the natural submersions \bea \rho_1\colon{\cal
W}\to  J^1\pi \ ,\ \rho_2\colon{\cal W}\to \Tan^*E \ ,\
\rho_{_{E}}\colon{\cal W}\to E \ ,\ \rho_{_{\Real}}\colon{\cal W}\to
\Real \label{project}
\\
\rho_1^r\colon{\cal W}_r\to  J^1\pi \ ,\ \rho_2^r\colon{\cal W}_r\to
J^1\pi^* \ ,\ \rho_{_{E}}^r\colon{\cal W}_r\to E \ ,\
\rho_{_{\Real}}^r\colon{\cal W}_r\to \Real \, .\nonumber
\eea
  Note that $\pi^1\circ\rho_1=\tau^1\circ\mu\circ\rho_2=\rho_{_{E}}$.
In addition, for $\bar y\in J^1\pi$, and
${\bf p}\in\Tan^*E$, there is also the natural projection
$$
\begin{array}{ccccc}
\mu_{_{\cal W}}&\colon& {\cal W}& \to &{\cal W}_r\\
& & (\bar y,{\bf p}) & \mapsto & (\bar y,[{\bf p}])
\end{array}
$$
where $[{\bf p}]=\mu({\bf p})\in J^1\pi^*$.
The bundle ${\cal W}$ is endowed with the following canonical structures:

\begin{definition}
\ben
\item
The {\rm coupling $1$-form} in ${\cal W}$ is the
$\rho_{_{\Real}}$-semibasic $1$-form $\hat{\cal C}\in\df^1({\cal
W})$ defined as follows: for every
$w=(j^1\phi(t),\alpha)\in {\cal W}$ (that is, $\alpha\in
T^*_{\rho_{_E}(w)} E$) and  $V\in \Tan_w{\cal W}$, then
$$
\hat{\cal C}(V)=\alpha(\Tan_w(\phi\circ\rho_{_{\Real}})V)\; .
$$
\item
The {\rm canonical $1$-form}
$\Theta_{\cal W}\in\df^1({\cal W})$ is the $\rho_{_{E}}$-semibasic form defined by
$\Theta_{\cal W}=\rho_2^*\Theta$.

The {\rm canonical $2$-form} is
$\Omega_{\cal W}=-\d\Theta_{\cal W}=\rho_2^*\Omega\in\df^2({\cal W})$.
\een
\label{coupling}
\end{definition}

Being $\hat{\cal C}$ a $\rho_{_{\Real}}$-semibasic form,
there is $\hat C\in\Cinfty ({\cal W})$ such that $\hat{\cal C}=\hat C\d t$.
 Note also that $\Omega_{\cal W}$ is degenerate, its kernel being the $\rho_2$-vertical vectors;
then $({\cal W},\Omega_{\cal W} )$ is a presymplectic manifold.

The local expressions for $\Theta_{\cal W}$, $\Omega_{\cal W}$, and $\hat{\cal C}$ are
$$
\Theta_{\cal W} = p_i\d q^i+p\d t  \quad , \quad
  \Omega_{\cal W} = -\d p_i\wedge\d q^i-\d p\wedge\d t   \quad , \quad
\hat{\cal C}=(p+p_i v^i)\d t \, .
$$

Given a Lagrangian density $\Lden\in\df^1( J^1\pi)$, we denote
$\hat\Lden=\rho_1^*\Lden\in\df^1({\cal W})$, and we can write
$\hat\Lden=\hat  L\d t$, with $\hat
 L =\rho_1^* L \in\Cinfty ({\cal W})$. We define a {\sl Hamiltonian
submanifold}
$$
{\cal W}_0=\{ w\in{\cal W}\ | \ \hat\Lden(w)=\hat{\cal C}(w) \}\, .
$$
So, ${\cal W}_0$ is the submanifold of $\cal W$ defined by the
regular constraint function $\hat C-\hat  L =0$.
Observe that this function is globally defined in $\cal W$,
using the dynamical data and the geometry.
In local coordinates this constraint function is
\beq
p + p_i v^i-\hat  L (t,q^j,v^j)=0
\label{lligam}
\eeq
and its meaning will be clear when we apply this formalism
to Optimal Control problems (see Section \ref{ocpg}).
The natural imbedding is
$\jmath_0\colon{\cal W}_0\hookrightarrow{\cal W}$,
and we have the projections (submersions), see diagram (\ref{diag0}):
$$
\rho_1^0\colon{\cal W}_0\to  J^1\pi \ ,\
\rho_2^0\colon{\cal W}_0\to \Tan^*E \ ,\
\rho_{_{E}}^0\colon{\cal W}_0\to E \ ,\
\rho_{_{\Real}}^0\colon{\cal W}_0\to \Real
$$
which are the restrictions to ${\cal W}_0$ of the projections
(\ref{project}), and
 $$
\hat {\rho }_2^0 = \mu\circ \rho _2^0\colon{\cal W}_0 \to J^1\pi^* \, .
$$
Local coordinates in ${\cal W}_0$ are
$(t,q^i,v^i,p_i)$, and we have that
$$
\begin{array}{ccc}
\rho_1^0(t,q^i,v^i,p_i) = (t,q^i,v^i)  & , &
\jmath_0(t,q^i,v^i,p_i) = (t,q^i,v^i, L -p_iv^i ,p_i) \\
\hat\rho_2^0(t,q^i,v^i,p_i) = (t,q^i,p_i) & , &
\rho_2^0(t,q^i,v^i,p_i) =(t,q^i, L -p_iv^i ,p_i) \, .
\end{array}
$$

\begin{prop}
${\cal W}_0$ is a $1$-codimensional $\mu_{_{\cal W}}$-transverse
submanifold
of ${\cal W}$, which is diffeomorphic to ${\cal W}_r$.
\label{1}
\end{prop}
\proof
For every $(\bar y,{\bf p})\in{\cal W}_0$, we have
$ L (\bar y)\equiv\hat  L (\bar y,{\bf p})=\hat C(\bar y,{\bf p})$, and
$$
(\mu_{_{\cal W}}\circ\jmath_0)(\bar y,{\bf p})= \mu_{_{\cal W}}(\bar
y,{\bf p})=(\bar y,\mu({\bf p}))\, .
$$

First, $\mu_{_{\cal W}}\circ\jmath_0$ is injective:
let $(\bar y_1,{\bf p}_1),(\bar y_2,{\bf p}_2)\in{\cal W}_0$, then we have
$$
(\mu_{_{\cal W}}\circ\jmath_0)(\bar y_1,{\bf p}_1)=
(\mu_{_{\cal W}}\circ\jmath_0)(\bar y_2,{\bf p}_2)
\,\Rightarrow\, (\bar y_1,\mu({\bf p}_1))=(\bar y_2,\mu({\bf p}_2))
\, \Rightarrow\,
\bar y_1=\bar y_2\ ,\ \mu({\bf p}_1)=\mu({\bf p}_2)
$$
hence
$ L (\bar y_1)= L (\bar y_2)=\hat C(\bar y_1,{\bf p}_1)=
\hat C(\bar y_2,{\bf p}_2)$.
In a local chart, the third equality gives
$$
p({\bf p}_1)+p_i({\bf p}_1)v^i(\bar y_1)=
p({\bf p}_2)+p_i({\bf p}_2)v^i(\bar y_2)
$$
but $\mu({\bf p}_1)=\mu({\bf p}_2)$ implies that
$$
p_i({\bf p}_1)=p_i([{\bf p}_1])=p_i([{\bf p}_2])=p_i({\bf p}_2)
$$
therefore $p({\bf p}_1)=p({\bf p}_2)$ and hence ${\bf p}_1={\bf p}_2$.

Second, $\mu_{_{\cal W}}\circ\jmath_0$ is onto, then, if $(\bar y,[{\bf
p}])\in{\cal W}_r$, there exists $(\bar y,{\bf
q})\in\jmath_0({\cal W}_0)$ such that $[{\bf q}]=[{\bf p}]$. In
fact, it suffices to take $[{\bf q}]$ such that, in a
local chart of $ J^1\pi \times_E\Tan^*E={\cal W}$
$$
p_i({\bf q})=p_i([{\bf p}]) \ , \ p({\bf q})= L (\bar y)-p_i([{\bf
p}])v^i(\bar y) \, .
$$

Finally, since ${\cal W}_0$ is defined by the constraint function
$\hat C-\hat  L $ and, as \(\displaystyle\ker\,\mu_{\cal W*}=\left\{\derpar{}{p}\right\}\)
 locally and
\(\displaystyle\derpar{}{p}(\hat C-\hat  L )=1\), then ${\cal W}_0$
is $\mu_{_{\cal W}}$-transversal.
 \qed

As a consequence of this result, the
submanifold ${\cal W}_0$ induces a section of the projection $\mu_{_{\cal W}}$,
$$
\hat h\colon{\cal W}_r\to{\cal W}\ .
$$
Locally, $\hat h$ is specified by giving the local {\sl
Hamiltonian function} $\hat H= -\hat  L +p_i v^i$; that is, $\hat
h(t,q^i,v^i,p_i)=(t,q^i,v^i,-\hat H,p_i)$. In this sense, $\hat h$
is said to be a {\sl Hamiltonian section} of $\mu_{_{\cal W}}$.

So we have the following diagram
\beq\bfig\xymatrix{&& J^1\pi &&\\
{\cal W}_0 \ar[urr]^{\txt{\small{$\rho_1^0$}}}
\ar[rr]^{\txt{\small{$\jmath_0$}}}
\ar[drr]^{\txt{\small{$\rho_2^0$}}}
\ar[ddrr]_{\txt{\small{$\hat\rho_2^0$}}}&& {\cal W}
\ar[u]^{\txt{\small{$\rho_1$}}}
\ar[d]_{\txt{\small{$\rho_2$}}}\ar[rr]^{\txt{\small{$\mu_{_{\cal
W}}$}}} && {\cal W}_r \ar[ull]_{\txt{\small{$\rho_1^r$}}}
\ar[dll]_{\txt{\small{$\rho_2\circ\hat h$}}}
\ar[ddll]^{\txt{\small{$\rho_2^r$}}}\\ && T^*E
\ar[d]_{\txt{\small{$\mu$}}}&& \\ && J^1\pi^* && }\efig
\label{diag0} \eeq

\section{Optimal control theory}
\protect\label{section-control}

\subsection{Classical formulation of Pontryagin's Maximum Principle}

In this section we consider non-autonomous optimal control
systems. This class of systems are determined by the {\sl state
equations}, which are a set of differential equations
\begin{equation}\label{general}
\dot{q}^i={\cal F}^i(t,q^j(t),u^a(t)) \,,\; 1 \le i \le n \, ,
\end{equation}
where $t$ is time,  $q^j$ denote the state
variables and  $u^a$, $1\leq a\leq m$, the
control inputs of the system that must be
determined. Prescribing initial conditions of the
state variables and fixing control inputs we
know completely the trajectory of the state
variables $q^j(t)$ (in the sequel, all the
functions are assumed to be at least $C^2$).
The objective is the following:

\begin{state}
\label{NAOCP} {\rm (Non-autonomous optimal control problem)}. Find
a $C^2$-piecewise smooth curve $\gamma(t)=(t,q^j(t),u^a(t))$ and
$T\in\R^+$ satisfying the conditions for the state variables at
time $0$ and $T$, the control equations (\ref{general}); and
minimizing the functional ${\mathcal J}(\gamma) = \int^T_0 \costf
(t,q^j(t),u^a(t))\,\d t \, .$
\end{state}

The solutions to this problem are called {\sl optimal trajectories}.

The necessary conditions to obtain the solutions to such a problem
are provided by {\sl Pontryagin's Maximum Principle} for
non-autonomous systems.  In this case, considering the time as another state variable,
we have \cite{P62}:

\begin{teor}
\label{pontry} {\rm (Pontryagin's Maximum Principle)}. If a
curve $\gamma: [0,T]\rightarrow \Real\times\Real^n\times\Real^m$,
$\gamma(t)=(t,q^i(t),u^a(t))$, with $\gamma(0)$ and
$\gamma(T)$ fixed, is an optimal trajectory, then there
exist functions $p(t)$, $p_i(t)$, $1\leq i\leq n$,
verifying:
 \bea \frac{d q^i}{d t}&=&
\displaystyle{\frac{\partial{\cal H}}{\partial
p_i}(t,q^i(t),u^a(t),p(t),p_i(t))}
\label{eqH1} \\
\frac{d p_i}{d t}&=& -\displaystyle{\frac{\partial{\cal
H}}{\partial q^i}(t,q^i(t),u^a(t),p(t),p_i(t))}
\label{eqH2} \\
{\cal
H}(t,q^i(t),u^a(t),p(t),p_i(t))&=&\underset{u^a}{\hbox{max}}\;
{\cal H}(t,q^i(t),u^a,p(t),p_i(t)), \quad t\in [0, T]
\label{eqH}
  \eea
and, moreover,
\beq
{\cal H}(t,q^i(t),u^a(t),p(t),p_i(t))=0, \quad t\in [0,T] \ ,
 \label{eqH0}
\eeq
 where
$$
{\cal H}(t,q^i,u^a,p,p_i)=p+p_j {\cal
F}^j(t,q^i,u^a)+p_0\costf (t,q^i,u^a)
$$
and $p_0\in\{-1,0\}$.
\end{teor}

When we are looking for extremal trajectories, which are those
satisfying the necessary conditions of Theorem \ref{pontry},
condition (\ref{eqH}) is usually replaced by the weaker condition
\begin{equation}\label{eqHweak}
 \varphi_a\equiv\frac{\partial{\cal H}}{\partial u^a}=0,
\quad 1\leq a\leq m\; .
\end{equation}
In this weaker form, the Maximum Principle only applies to
optimal trajectories with optimal controls interior to the
control set.

{\bf Remark}: An extremal trajectory is called {\sl normal} if
$p_0=-1$ and {\sl abnormal} if $p_0=0$. For the sake of
simplicity, we only consider normal extremal trajectories, but the
necessary conditions for abnormal extremals can also be
characterized geometrically using the formalism given in Section
\ref{srfnas}. Hence, from now on we will take $p_0=-1$.

An optimal control problem is said to be {\sl regular} if the
following matrix has maximal rank
\beq \left(\frac{\partial \varphi_a}{\partial u^b}\right) =
\left(\frac{\partial^2 {\cal H}}{\partial u^a\partial u^b}\right) \ .
\label{31} \eeq

In the following sections we develop a geometric
formulation of this Maximum Principle in its weak form,
similar to the Skinner-Rusk approach to non-autonomous
mechanics as was explained in Section \ref{srfnas} and
references therein.

\subsection{Unified geometric framework for optimal control theory}
\protect\label{ocpg}

In a global description, we have a fiber
bundle structure $\pi^C\colon C \longrightarrow E$
and $\pi\colon E\to \R$, where $E$ is equipped
with natural coordinates $(t,q^i)$ and $C$ is
the bundle of controls, with coordinates  $(t,q^i,u^a)$.

The state equations can be geometrically
described as a smooth map
${\cal F}: C\longrightarrow J^1\pi$
such that it makes commutative the following diagram
$$
\bfig\xymatrix{C \ar[rrrr]^{\txt{\small{${\cal F}$}}}
\ar[drr]^{\txt{\small{$\pi^C$}}}
\ar[ddrr]_{\txt{\small{$\bar\pi^C$}}}&&&& J^1\pi
\ar[dll]_{\txt{\small{$\pi^1$}}}\ar[ddll]^{\txt{\small{$\bar\pi^1$}}}
\\ && E \ar[d]_{\txt{\small{$\pi$}}}
&& \\ && \Real && }\efig
$$
which means that ${\cal F}$ is a jet
field along $\pi^C$ and also along $\bar\pi^C$. Locally we have
${\cal F}(t,q^i,u^a)=(t,q^i,{\cal F}^i(t,q^i,u^a))$.

Geometrically, we will assume that an {\sl optimal control
system} is determined by the pair $(\Lag, {\cal F})$, where
$\Lag\in\df^1(C)$ is a $\bar{\pi}^C$-semibasic $1$-form, then
$\Lag=\costf \d t$, with $\costf \in\Cinfty(C)$ representing the
cost function; and ${\cal F}$ is the jet field introduced in the
above section.

In this framework, Theorem \ref{pontry} in its weak form
can be restated as:
\begin{teor}
\label{pontry2} If a curve $\gamma: I\rightarrow  C$, with
$\gamma(0)$ and $\gamma(T)$ fixed, is an optimal
trajectory, then there exists a curve $\Gamma: I\rightarrow
C\times_E T^*E$ such that, in a natural coordinate system,
$\Gamma(t)=(\gamma(t), p(t),p_i(t))$ verifies (\ref{eqH1}),
(\ref{eqH2}), (\ref{eqH0}) and (\ref{eqHweak}), where ${\cal
H}=p+p_j {\cal F}^j+p_0\costf $ and $p_0\in\{-1,0\}$.
\end{teor}

Now, we develop the geometric model of Optimal Control theory
according to the Skinner-Rusk formulation.

The graph of the mapping ${\cal F}$, $\hbox{Graph}\, {\cal
F}$, is a subset of $C\times_E J^1\pi$ and allows us to
define the {\sl extended} and the {\sl restricted control-jet-momentum bundles},
respectively:
$$
\mathcal{W}^{\cal F}=\hbox{Graph}\, {\cal F}\times_E T^*E
\quad , \quad
\mathcal{W}^{\cal F}_{r}=\hbox{Graph}\, {\cal F}\times_E
J^1\pi^*
$$
which are submanifolds of $C\times_E{\mathcal W}=C\times_EJ^1\pi\times_{E}\Tan^*E$
and $C\times_E{\mathcal W}_r=C\times_EJ^1\pi\times_{E}J^1\pi^*$, respectively.

In $\mathcal{W}^{{\cal F}}$ and $\mathcal{W}^{\cal F}_{r}$ we
have natural coordinates $(t,q^i,u^a,p,p_i)$
and $(t,q^i,u^a,p_i)$, respectively.
We have the immersions (see diagram (\ref{diag00})):
\beann
i^{{\cal F}}&\colon& \mathcal{W}^{{\cal F}}\hookrightarrow
C\times_E \mathcal{W}\, , \quad
i^{\cal F}(t,q^i,u^a,p,p_i)=(t,q^i,u^a,{\cal F}^i(t,q^j,u^b,),p,p_i)\\
i^{\cal F}_{r}&\colon& \mathcal{W}^{\cal F}_{r}\hookrightarrow
C\times_E \mathcal{W}_r\, , \quad
i^{\cal F}_{r}(t,q^i,u^a,p_i)=(t,q^i,u^a,{\cal F}^i(t,q^j,u^b),p_i) \ ,
\eeann
and taking the natural projection
$$
\sigma_{\cal W}\colon C\times_E{\mathcal W}\to {\mathcal W}
$$
we can construct the pullback of the
coupling 1-form $\hat{\mathcal C}$ and
of the forms $\Theta_{\mathcal W}$
and $\Omega_{\mathcal W}$ to ${\mathcal W}^{\cal F}$:
\[
{\mathcal C}_{\mathcal{W}^{\cal F}}= (\sigma_{\cal W}\circ i^{\cal
F})^*\hat{\mathcal C} \quad , \quad \Theta_{{\mathcal W}^{\cal
F}}=(\sigma_{\cal W}\circ i^{\cal F})^*\Theta_{\mathcal W} \quad ,
\quad \Omega_{{\mathcal W}^{\cal F}}=(\sigma_{\cal W}\circ i^{\cal
F})^*\Omega_{\mathcal W}= (\rho_2^{\cal F})^*\Omega \, ,
\]
see Definition \ref{coupling}, whose local expressions are:
$$
{\mathcal C}_{\mathcal{W}^{\cal F}}=(p+p_i {\cal
F}^i(t,q^j,u^a))\d t\quad , \quad 
\Theta_{{\mathcal W}^{\cal F}} = p_i\d q^i+p\d t \quad , \quad
\Omega_{{\mathcal W}^{\cal F}} = -\d p_i\wedge\d q^i-\d p\wedge\d t
 \, .
$$
Hence, we can draw the diagram
\beq
\bfig\xymatrix{
 C\times_E{\cal W}
 \ar[rrrrrr]^{\txt{\small{${\rm Id}\times\mu_{\cal W}$}}}
\ar[rrdddd]_{\txt{\small{$\sigma_{\cal W}$}}} & & & & & &
 C\times_E{\cal W}_r
 \ar[lldddd]^{\txt{\small{$\sigma_{{\cal W}_r}$}}} \\
& & & {\rm Graph}\,{\cal F} & & &
 \\
 & & {\cal W}^{\cal F}\ar[ru]
 \ar[rr]^{\txt{\small{$\mu_{{\cal W}^{\cal F}}$}}}
\ar[lluu]_{\txt{\small{$i^{\cal F}$}}}
\ar[rd]_{\txt{\small{$\rho_2^{\cal F}$}}}
 & & {\cal W}_r^{\cal F}\ar[lu]
  \ar[rruu]^{\txt{\small{$i^{\cal F}_r$}}}& & \\
 & & & \Tan^*E & & &
 \\
  & & {\cal W}
\ar[ur]^{\txt{\small{$\rho_2$}}}
\ar[rr]^{\txt{\small{$\mu_{\cal W}$}}}
& &
 {\cal W}_r & &
}\efig
\label{diag00}
\eeq
where $\rho_2^{\cal F}$, $\rho_2$, $\mu_{{\cal W}^{\cal F}}$, and $\sigma_{{\cal W}_r}$
are natural projections.

Furthermore we can define the unique function
$H_{{\mathcal W}^{\cal F}}: {\mathcal W}^{\cal F}\longrightarrow \R$
by the condition
$$
{\mathcal C}_{\mathcal{W}^{\cal F}}-(\rho^{\cal
F}_{1})^*\Lag=H_{{\mathcal W}^{\cal F}}\d t \, .
$$
where $\rho^{\cal F}_{1}\colon {\mathcal W}^{\cal F}\to C$
is another natural projection.
This function $H_{{\mathcal W}^{\cal F}}$ is locally described as
\beq
H_{{\mathcal W}^{\cal F}}(t,q^i,u^a,p,p_i)=
p+p_i{\cal F}^i(t,q^j,u^a)-\costf (t,q^j,u^a) \, ;
\label{phf}
\eeq
(compare this expression with (\ref{lligam})).
This is the natural Pontryagin Hamiltonian function
as appears in Theorem \ref{pontry}.

Let ${\mathcal W}^{\cal F}_0$ be the submanifold of
$\mathcal{W}^{\cal F}$ defined by the vanishing of
$H_{{\mathcal W}^{\cal F}}$; that is,
\[
{\mathcal W}^{\cal F}_0= \{ w\in {\mathcal W}^{\cal F} \ | \
H_{{\mathcal W}^{\cal F}}(w)=0\} \, .
\]
In local coordinates, ${\mathcal W}^{\cal F}_0$ is given by the constraint
\[
p+p_i {\cal F}^i(t,q^j,u^a)-\costf (t,q^j,u^a)=0 \ .
\]
Observe that, in this way, we recover the condition (\ref{eqH0}).
 An obvious set of coordinates in ${\mathcal W}^{\cal F}_0$ is
$(t,q^i,u^a,p_i)$.
We denote by
$\jmath_0^{\cal F}\colon{\mathcal W}^{\cal F}_0 \to {\mathcal W}^{\cal F}$
the natural embedding; in local coordinates,
\[
\jmath_0^{\cal F}(t,q^i,u^a,p_i)=
(t,q^i,u^a,\costf (t,q^j,u^b)-p_i {\cal F}^i(t,q^j,u^b),p_j) \ .
\]

In a similar way to Proposition \ref{1}, we may prove the following:

\begin{prop}
\label{prop2}
${\cal W}^{\cal F}_0$ is a $1$-codimensional
$\mu_{_{{\cal W}^{\cal F}}}$-transverse submanifold
of ${\cal W}^{\cal F}$, diffeomorphic to ${\cal W}^{\cal F}_r$.
\end{prop}

As a consequence, the submanifold ${\cal W}^{\cal F}_0$ induces a
section of the projection $\mu_{_{{\cal W}^{\cal F}}}$, \beq \hat
h^{\cal F}\colon{\cal W}^{\cal F}_r\to{\cal W}^{\cal F}\ .
\label{hamsect} \eeq Locally, $\hat h^{\cal F}$ is specified by
giving the local {\sl Hamiltonian function} $\hat H^{\cal
F}=p_j{\cal F}^j-\costf $; that is, $\hat h^{\cal
F}(t,q^i,u^a,p_i)=(t,q^i,u^a,p=-\hat H^{\cal F},p_i)$. The map
$\hat h^{\cal F}$ is called a {\sl Hamiltonian section} of
$\mu_{_{{\cal W}^{\cal F}}}$.

Thus, we can draw the diagram, where all the projections are natural
\beq
\bfig\xymatrix{& &  J^1\pi
 \ar[rrdd]^{\txt{\small{$\bar\pi^1$}}}
\ar[lldd]_{\txt{\small{$\pi^1$}}} & & \\ \\
E & &  C \ar[uu]^{\txt{\small{${\cal F}$}}}
 \ar[rr]^{\txt{\small{$\bar\pi^C$}}}
 \ar[ll]_{\txt{\small{$\pi^C$}}} & &  \Real \\ \\
 {\cal W}^{\cal F}_0
\ar[uurr]^(.7){\txt{\small{$\rho_1^{0{\cal F}}$}}}
\ar[rr]^{\txt{\small{$\jmath_0^{\cal F}$}}}
\ar[ddrr]^{\txt{\small{$\rho_2^{0{\cal F}}$}}}
\ar[ddddrr]_{\txt{\small{$\hat\rho_2^{0{\cal F}}$}}}
\ar[uu]^{\txt{\small{$\rho_{_E}^{0{\cal F}}$}}} &&
 {\cal W}^{\cal F} \ar[uu]^{\txt{\small{$\rho_1^{\cal F}$}}}
\ar[dd]_{\txt{\small{$\rho_2^{{\cal F}}$}}}
\ar[rr]^{\txt{\small{$\mu_{_{{\cal W}^{\cal F}}}$}}}
\ar[uurr]^(.3){\txt{\small{$\rho_{_\Real}^{\cal F}$}}}
\ar[uull]^(.7){\txt{\small{$\rho_{_E}^{\cal F}$}}} &&
 {\cal W}^{\cal F}_r \ar[uull]_(.3){\txt{\small{$\rho_1^{r{\cal F}}$}}}
 \ar[ddll]_(0.4){\txt{\small{$\rho_2^{r{\cal F}}\circ\hat h^{\cal
F}$}}} \ar[ddddll]^{\txt{\small{$\rho_2^{r{\cal F}}$}}}
\ar[uu]_{\txt{\small{$\rho_{_\Real}^{r{\cal F}}$}}} \\ \\
&& T^*E\ar[dd]_{\txt{\small{$\mu$}}} &&\\ \\ && J^1\pi^* &&
}\efig
\label{diag000}
\eeq

Finally we define the forms
$$
\Theta_{{\mathcal W}_0^{\cal F}}=
(\jmath_0^{\cal F})^*\Theta_{{\mathcal W}^{\cal F}} \quad , \quad
\Omega_{{\mathcal W}_0^{\cal F}}=(\jmath_0^{\cal F})^*\Omega_{{\mathcal
W}^{\cal F}}
$$
with local expressions
$$
\Theta_{{\mathcal W}_0^{\cal F}}=p_i\d q^i+(\costf -p_i{\cal F}^i)\d
t \quad , \quad \Omega_{{\mathcal W}_0^{\cal F}}=-\d p_i\wedge\d
q^i-\d (\costf -p_i{\cal F}^i)\wedge\d t \, .
$$

\subsection{Optimal Control equations}
\protect\label{oce}

Now we are going to establish the dynamical problem for the
system $({\cal W}_0^{\cal F},\Omega_{{\mathcal W}_0^{\cal
F}})$ and as a consequence we obtain a  geometrical version
of the weak form of the Maximum Principle.

\begin{prop} Let $(\Lag,{\cal
F})$ define a regular optimal control problem, then there
exists a submanifold ${\mathcal W}^{\cal F}_1$ of
${\mathcal W}^{\cal F}_0$ and a
unique vector field $Z\in\vf({\cal W}_0^{\cal F})$
tangent to ${\mathcal W}^{\cal F}_1$ such that
\begin{equation}
\label{vhs}
[\inn(Z)\Omega_{{\mathcal W}_0^{\cal F}}]\vert_{{\mathcal W}^{\cal F}_1}=0 \quad
, \quad [\inn(Z)\d t]\vert_{{\mathcal W}^{\cal F}_1}=1\ .
\end{equation}
The integral curves $\Gamma$ of $Z$ satisfy locally the
necessary conditions of Theorem \ref{pontry2}.
\end{prop}

\proof In a natural coordinate system, we have
\[
Z=f\frac{\partial }{\partial t}+A^i\frac{\partial}{\partial q^i}+
B^a\frac{\partial}{\partial u^a}+C_i\frac{\partial}{\partial p_i}
\]
where $f,A^i, B^a, C_i$ are unknown functions in ${\mathcal W}_0^{\cal F}$.
Then, the second equation (\ref{vhs}) leads to $f=1$, and from the first
we obtain that
\bea
&\mbox{\rm coefficients in $\d p_i$} :& {\cal F}^i-A^i=0
\label{22} \\
& \mbox{\rm coefficients in $\d u^a$} :&
\frac{\partial \costf }{\partial u^a}-p_j\frac{\partial {\cal F}^j}{\partial
u^a}=0
\label{21} \\
& \mbox{\rm coefficients in $\d q^i$} :&
\frac{\partial \costf }{\partial q^i}-p_j\frac{\partial {\cal F}^j}{\partial
q^i}-C_i=0
\label{20} \\
& \mbox{\rm coefficients in $\d t$} \ :& -A^i\frac{\partial
\costf }{\partial q^i} +A^ip_j\frac{\partial {\cal
F}^j}{\partial q^i}- B^a\frac{\partial \costf }{\partial
u^a}+ B^ap_j\frac{\partial {\cal F}^j}{\partial
u^a}+C_i{\cal F}^i=0 \, . \label{19} \eea Now, if
$\Gamma(t)=(t,q^i(t),u^a(t),p_i(t))$ is an integral curve
of $Z$, we have that \(\displaystyle A^i=\frac{d q^i}{d
t}\), \(\displaystyle B^a=\frac{d u^a}{d t}\),
\(\displaystyle C_i=\frac{d p_i}{d t}\).

The {\sl Pontryagin Hamiltonian function} is ${\cal
H}=p+p_i {\cal F}^i-\costf $. As we are in ${\mathcal
W}^{\cal F}_0$, condition (\ref{eqH0}), ${\cal H}=0$,  is
satisfied. Furthermore, \bit
\item
   From  (\ref{22}) we deduce that $A^i={\cal F}^i$; that is,
\(\displaystyle \frac{d q^i}{d t}=\frac{\partial{\cal H}}{\partial
p_i}\), which are the equations (\ref{eqH1}).
\item
Equations (\ref{21}) determine a new set of conditions
\beq
\varphi_a=
\frac{\partial \costf }{\partial u^a}-p_j\frac{\partial {\cal F}^j}{\partial u^a}=
\frac{\partial{\cal H}}{\partial u^a}=0
\label{veinte}
\eeq
which are equations (\ref{eqHweak}). We assume that they
define the new submanifold ${\mathcal W}^{\cal F}_1$ of
${\mathcal W}^{\cal F}_0$. We denote by $\jmath_1^{\cal
F}\colon{\cal W}_1^{\cal F}\hookrightarrow{\cal W}_0^{\cal
F}$ the natural embedding.
\item
{}From  (\ref{20}) we completely determine the functions
\(\displaystyle C_i=\frac{d p_i}{d t}=-\derpar{{\cal H}}{q^i}\);
which are the equations (\ref{eqH2}).
\item
Finally, using (\ref{22}), (\ref{20}) and (\ref{21}) it is
easy to prove that equations (\ref{19}) hold identically.
\eit
Furthermore $Z$ must be tangent to ${\cal W}_1^{\cal F}$, that is,
\[
Z(\varphi_a)=Z\left(\frac{\partial{\cal H}}{\partial u^a}\right)=0
\qquad \mbox{\rm (on ${\cal W}_1^{\cal F}$)}
\]
or, in other words, \beq 0=\frac{\partial^2 {\cal H}}{\partial t
\partial u^a}+ {\cal F}^i\frac{\partial^2 {\cal H}}{\partial
q^i\partial u^a}+ B^b\frac{\partial^2 {\cal H}}{\partial
u^b\partial u^a} - \frac{\partial {\cal H}}{\partial q^i}
\frac{\partial^2 {\cal H}}{\partial p_i\partial u^a} \qquad
\mbox{\rm (on ${\cal W}_1^{\cal F}$)} \, . \label{tang}
\eeq However, as the optimal control problem is regular,
the matrix $\displaystyle{\frac{\partial^2 {\cal
H}}{\partial u^b\partial u^a}}$ has maximal rank. Then the
equations (\ref{tang}) determine all the coefficients
$B^b$. \qed

As a direct consequence of this proposition, we state the
intrinsic version of Theorem \ref{pontry2}.

\begin{teor}
{\rm (Geometric weak Pontryagin's Maximum Principle)}. If
$\gamma\colon I \rightarrow C$ is a solution to the regular
optimal control problem given by $(\Lag,{\cal F})$, then
there exists an integral curve of a vector field
$Z\in\vf({\cal W}_0^{\cal F})$,  whose projection to $C$ is
$\gamma$, and such that $Z$ is a solution to the equations
$$
\inn(Z)\Omega_{{\mathcal W}_0^{\cal F}}=0 \quad
, \quad \inn(Z)\d t=1\ ,
$$
in a submanifold ${\cal W}_1^{\cal F}$ of ${\cal W}_0^{\cal F}$, which is given
by the condition (\ref{veinte}).
\end{teor}

Note that the conditions fulfilled by the integral curves
of $Z$, satisfying the suitable initial
conditions, imply that their natural projections on $C$ are
$\gamma$.

 {\bf Remark}: In fact, the second
equation of (\ref{vhs}) could be relaxed to the condition
$$
\inn(Z)\d t\not=0 \ ,
$$
which determines vector fields transversal to $\pi$ whose integral curves
are equivalent to those obtained above, with arbitrary reparametrization.

Note that, using  the implicit
function theorem on the equations $\varphi_a=0$, we
get the functions $u^a=u^a(t,q,p)$. Therefore, for regular
control problems, we can choose local coordinates
$(t,q^i,p_i)$ on ${\mathcal W}^{\cal F}_1$, and
${\cal H}|_{{\mathcal W}^{\cal F}_1}$ is locally a function of
these coordinates.

If the control problem is not regular, then
one has to implement a constraint algorithm to
obtain a final constraint submanifold
${\mathcal W}^{\cal F}_f$ (if it exists) where the vector field
$Z$ is tangent (see, for instance, \cite{DI-2003}).

Let $\jmath_1\colon {\mathcal W}_1^{\cal F}\rightarrow {\mathcal
W}_0^{\cal F}$ be the natural embedding, the form $\Omega_{{\mathcal
W}_1^{\cal F}}= (\jmath_1^{\cal F})^*\Omega_{{\mathcal W}_0^{\cal
F}}$ is locally written as
\[
\Omega_{{\mathcal W}_1^{\cal F}}=
 - \d p_i\wedge\d q^i-\d {\cal H}|_{{\mathcal W}^{\cal F}_1}\wedge\d t \, .
\]
Hence, for optimal control problems,
taking into account the regularity of the matrix (\ref{31}),
we have the following:

\begin{prop}\label{cosympl}
If the optimal control problem is regular, then $({\mathcal
W}_1^{\cal F},\Omega_{{\mathcal W}_1^{\cal F}},\d t)$ is a
cosymplectic manifold, that is, $(\Omega_{{\mathcal
W}_1^{\cal F}})^{n} \wedge \d t$ is a volume form
(see \cite{LR}).
\end{prop}

\section{Implicit optimal control problems}
\protect\label{iocp}

\subsection{Unified geometric framework for implicit optimal control problems}

The formalism presented in  Section \ref{ocpg} is valid for a more
general class of optimal control problems not previously
considered from a geometric perspective: optimal control problems
whose state equations are {\sl implicit}, that is,
\begin{equation}
\label{impli} \Psi^{\alpha}(t,q,\dot{q},u)=0\ ,\ 1\leq \alpha\leq s\
,\ \mbox{\rm with $\d\Psi^1\wedge\ldots\wedge\d\Psi^s\not=0$} \, .
\end{equation}
There are several examples of these kinds of optimal control
problems, some of them coming from engineering
applications. In Section \ref{examples} we study two
specific examples: the descriptor systems which appear in
electrical engineering and the controlled Lagrangian
systems which play a relevant role in robotics.

From a  more geometric point of view, we may interpret
Equations (\ref{impli}) as constraint functions determining
a submanifold $M_C$ of $C\times_E J^1\pi$, with natural
embedding $\jmath^{M_C}\colon M_C\hookrightarrow C\times_E
J^1\pi$. We will also assume that
$(\pi^C\times\pi^1)\circ\jmath^{M_C}\colon M_C\to E$ is a
surjective submersion.

In this situation, the techniques presented in the previous
section are still valid. Now the implicit optimal control system
is determined by the data $(\Lag, M_C)$, where $\Lag\in\df^1(M_C)$
is a semibasic form with respect to the projection
$\tau^{M_C}\colon M_C\to\Real$, and hence it can be written as
$\Lag= \costf \d t$, for some $\costf \in\Cinfty (M_C)$. First
define  the {\sl extended control-jet-momentum manifold}
 and the {\sl restricted control-jet-momentum manifold}
$$
\mathcal{W}^{M_C}=M_C\times_E T^*E\quad ,\quad
\mathcal{W}^{M_C}_{r}=M_C\times_E J^1\pi^*
$$
which are submanifolds of $C\times_E{\mathcal W}=C\times_EJ^1\pi\times_{E}\Tan^*E$
and $C\times_E{\mathcal W}_r=C\times_EJ^1\pi\times_{E}J^1\pi^*$, respectively.

We have the canonical immersions (embeddings)
$$
i^{M_C}\colon \mathcal{W}^{{M_C}}\hookrightarrow C\times_E{\mathcal
W}\quad ,\quad i^{M_C}_r\colon \mathcal{W}^{M_C}_{r}\hookrightarrow
C\times_E{\mathcal W}_r\ .
$$
So we can draw a diagram analogous to (\ref{diag00}) replacing the
core of the diagram by
$$
 \bfig\xymatrix{ & M_C &
\\
 {\cal W}^{M_C}
\ar[ru]^{\txt{\small{$\rho^{M_C}_1$}}}
\ar[rr]^{\txt{\small{$\mu_{{\cal W}^{M_C}}$}}}
 & & {\cal W}_r^{M_C}
\ar[lu]_{\txt{\small{$\rho^{rM_C}_1$}}}
 }\efig
 $$
 where all the projections are natural.

Now, consider the pullback of the coupling 1-form $\hat{\mathcal C}$
and the forms $\sigma_{\cal W}^*\Theta_{\mathcal W}$ and
$\sigma_{\cal W}^*\Omega_{\mathcal W}$ to ${\mathcal W}^{M_C}$
by the map
$i^{M_C}\colon {\mathcal{W}^{M_C}} \to C\times_E{\mathcal{W}}$; that is
\[
{\mathcal C}_{\mathcal{W}^{M_C}}=(\sigma_{\cal W}\circ i^{M_C})^*\hat{\mathcal C} \ ,\
\Theta_{{\mathcal W}^{M_C}}=(\sigma_{\cal W}\circ i^{M_C})^*\Theta_{\mathcal W} \ , \
\Omega_{{\mathcal W}^{M_C}}=(\sigma_{\cal W}\circ i^{M_C})^*\Omega_{\mathcal W}\ ,
\]
and denote by $\hat C\in\Cinfty ({\mathcal{W}^{M_C}})$ the unique function such that
${\mathcal C}_{\mathcal{W}^{M_C}}=\hat C\d t$.
Finally, let  $H_{{\mathcal W}^{M_C}}: {\mathcal W}^{M_C}\to \R$ be the
unique function such that
${\mathcal C}_{\mathcal{W}^{M_C}}-(\rho^{M_C}_1)^*\Lag=H_{{\mathcal W}^{M_C}}\d t$.
Observe that
$H_{{\mathcal W}^{M_C}}=\hat C-\hat \costf $, where $\hat \costf =(\rho^{M_C}_1)^*\costf $,
and remember that $H_{{\mathcal W}^{M_C}}$ is the Pontryagin Hamiltonian function,
see (\ref{phf}).

Let ${\mathcal W}^{M_C}_0$ be the submanifold of
$\mathcal{W}^{M_C}$  defined by the vanishing of
$H_{{\mathcal W}^{M_C}}$, i.e.
\beq
{\mathcal W}^{M_C}_0=\{ w\in {\mathcal W}^{M_C} \; |\;
H_{{\mathcal W}^{M_C}}(w)=(\hat C-\hat \costf )(w)=0\} \ ,
\label{noseque}
\eeq
and denote by
$\jmath_0^{M_C}:  {\mathcal W}^{M_C}_0 \hookrightarrow{\mathcal W}^{M_C}$
the natural embedding.
As in Proposition \ref{1} we may prove the following:

\begin{prop}\label{prop2-b}
${\cal W}^{M_C}_0$ is a $1$-codimensional $\mu_{{\mathcal
W}^{M_C}}$-transverse submanifold of ${\cal W}^{M_C}$,
diffeomorphic to ${\cal W}^{M_C}_r$.
\end{prop}

As a consequence, the
submanifold ${\cal W}^{\cal F}_0$ induces a section of the projection $\mu_{_{{\cal W}^{M_C}}}$,
$$
\hat h^{M_C}\colon{\cal W}^{M_C}_r\to{\cal W}^{M_C}\ .
$$
Then we can draw the following diagram, which is analogous to (\ref{diag000}),
where all the projections are natural
$$
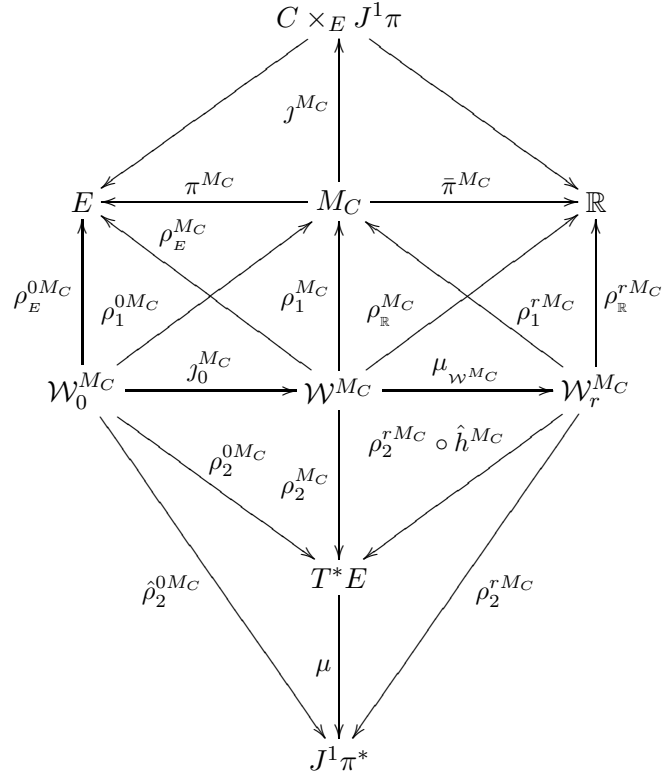
\bfig\xymatrix{& &  C\times_EJ^1\pi \ar[rrdd]^{\txt{\small{$ $}}}
\ar[lldd]_{\txt{\small{$ $}}} & & \\ \\
E & &  M_C \ar[uu]^{\txt{\small{$\jmath^{M_C}$}}}
\ar[rr]^{\txt{\small{$\bar\pi^{M_C}$}}}
\ar[ll]_{\txt{\small{$\pi^{M_C}$}}} & &  \Real \\ \\
{\cal W}^{M_C}_0
\ar[uurr]^(.3){\txt{\small{$\rho_1^{0{M_C}}$}}}
\ar[rr]^{\txt{\small{$\jmath_0^{M_C}$}}}
\ar[ddrr]^{\txt{\small{$\rho_2^{0{M_C}}$}}}
\ar[ddddrr]_{\txt{\small{$\hat\rho_2^{0{M_C}}$}}}
\ar[uu]^{\txt{\small{$\rho_{_E}^{0{M_C}}$}}} &&
{\cal W}^{M_C} \ar[uu]^{\txt{\small{$\rho_1^{M_C}$}}}
\ar[dd]_{\txt{\small{$\rho_2^{{M_C}}$}}}
\ar[rr]^{\txt{\small{$\mu_{_{{\cal W}^{M_C}}}$}}}
\ar[uurr]^(.3){\txt{\small{$\rho_{_\Real}^{M_C}$}}}
\ar[uull]_(.7){\txt{\small{$\rho_{_E}^{M_C}$}}} &&
{\cal W}^{M_C}_r \ar[uull]_(.3){\txt{\small{$\rho_1^{r{M_C}}$}}}
\ar[ddll]_(.4){\txt{\small{$\rho_2^{r{M_C}}\circ\hat h^{M_C}$}}}
\ar[ddddll]^{\txt{\small{$\rho_2^{r{M_C}}$}}}
\ar[uu]_{\txt{\small{$\rho_{_\Real}^{r{M_C}}$}}} \\ \\
&& T^*E\ar[dd]_{\txt{\small{$\mu$}}} &&\\ \\ && J^1\pi^* &&
}\efig
$$
Finally, we define the forms
$$
\Theta_{{\mathcal W}_0^{M_C}}=(\jmath_0^{M_C})^*\Theta_{{\mathcal
W}^{M_C}} \quad ,\quad \Omega_{{\mathcal
W}_0^{M_C}}=(\jmath_0^{M_C})^*\Omega_{{\mathcal W}^{M_C}} \ .
$$

\subsection{Optimal Control equations}
\protect\label{sub-imp}

Now, we will see how the dynamics of the optimal control problem
$(\Lag, {M_C})$ is determined by the solutions (where they exist)
of the equations
\begin{equation}\label{vhs1}
\inn(Z)\Omega_{{\mathcal W}_0^{M_C}}=0 \quad , \quad
\inn(Z)\d t =1
\quad , \quad
\mbox{\rm for $Z\in\vf({\cal W}_0^{M_C})$} \ .
\end{equation}
As in Section \ref{oce} , the second equation of (\ref{vhs1}) can be relaxed to the condition
$$
\inn(Z)\d t\not=0 \ .
$$

In order to work in local coordinates we need the
following proposition, whose proof is obvious:

\begin{prop}
For a given $w\in {\mathcal W}_0^{M_C}$, the
following conditions are equivalent:
\begin{enumerate}
\item  There exists  a vector $Z_w\in T_w {\mathcal W}_0^{M_C}$ verifying that
\[
\Omega_{{\mathcal W}_0^{M_C}}(Z_w, Y_w) =0\ , \ \mbox{\rm for every
$Y_w\in T_w {\mathcal W}_0^{M_C}$} \ .
\]
\item  There exists a vector $Z_w\in T_w (C\times_E {\mathcal W})$
verifying that
 \begin{enumerate}
\item[(i)] $Z_w\in T_w {\mathcal W}_0^{M_C}$,
\item[(ii)] $\inn({Z_w})(\sigma_{\cal W}^*\Omega_{\mathcal W})_w\in (T_w {\mathcal W}_0^{M_C})^0$
\ .
\end{enumerate}
\end{enumerate}
\label{2cons}
\end{prop}

As a consequence of this last proposition, we can obtain
the implicit optimal control equations using condition 2 as
follows: there exists $Z\in\vf(C\times_E{\mathcal W})$ such
that \bit
\item[(i)]
$Z$ is tangent to ${\mathcal W}_0^{M_C}$.
\item[(ii)]
The $1$-form $\inn(Z)\sigma_{\cal W}^*\Omega_{\mathcal W}$
is null on the vector fields tangent to ${\mathcal W}_0^{M_C}$.
\eit
As ${\mathcal W}_0^{M_C}$ is defined in (\ref{noseque}), and the constraints are
 $\Psi^\alpha=0$ and $\hat{C}-\hat{\costf }=0$; then there exist
$\lambda_{\alpha},\lambda\in\Cinfty(C\times_E{\mathcal W})$,
to be determined, such that
\[
(\inn({Z})\sigma_{\cal W}^*\Omega_{\mathcal W})\vert_{{\mathcal
W}_0^{M_C}}= (\lambda_{\alpha}\d\Psi^{\alpha} +\lambda
\d(\hat{C}-\hat{\costf }))\vert_{{\mathcal W}_0^{M_C}} \ .
\]
As usual, the undetermined functions $\lambda_{\alpha}$'s and $\lambda$
are called Lagrange multipliers.

Now using coordinates $(t, q^i, u^a, v^i, p, p^i)$ in $C\times_E
{\mathcal W}$, we look for a  vector field
\[
Z=\frac{\partial}{\partial t}+A^i\frac{\partial}{\partial q^i}+
B^a\frac{\partial}{\partial u^a}+C^i\frac{\partial}{\partial v^i}
+D_i\frac{\partial}{\partial p_i}+E\frac{\partial}{\partial p} \ ,
\]
where $A^i, B^a, C^i, D_i, E$ are unknown functions in
${\mathcal W}_0^{M_C}$ verifying the equation
\beann
0&=&i_Z\left(\d q^i\wedge \d p_i+\d t\wedge \d p\right)-\lambda_{\alpha} \d \Psi^{\alpha}
-\lambda \d (p+p_iv^i-\costf (t,q,u))\nonumber\\
&=& \left(-E-\lambda_{\alpha}\frac{\partial\Psi^{\alpha}}{\partial t}+
\lambda \frac{\partial \costf }{\partial t}\right)\d t
+\left(\lambda\frac{\partial \costf }{\partial q^i}-
\lambda_{\alpha}\frac{\partial \Psi^{\alpha}}{\partial q^i}-D_i\right)\d q^i\\
& & +\left(\lambda\frac{\partial \costf }{\partial u^a}
-\lambda_{\alpha}\frac{\partial\Psi^{\alpha}}{\partial u^a}\right)\d u^a
+\left(-\lambda p_i-\lambda_{\alpha}\frac{\partial \Psi^{\alpha}}{\partial v^i}\right)\d v^i\\
&& +(A^i-\lambda v^i)\d p_i+(1-\lambda)\d p \ .
\eeann
 Thus, we obtain $\lambda=1$, and
$$
A^i=v^i \ ,\
D_i=\displaystyle{\frac{\partial \costf }{\partial q^i}-
\lambda_{\alpha}\frac{\partial \Psi^{\alpha}}{\partial q^i}}\ ,\
E=\frac{\partial \costf }{\partial t}-\lambda_{\alpha}\frac{\partial \Psi^{\alpha}}{\partial t}\ ,\
p_i=-\lambda_{\alpha} \frac{\partial \Psi^{\alpha}}{\partial v^i}\ ,\
0=\frac{\partial \costf }{\partial u^a}-\lambda_{\alpha}\frac{\partial\Psi^{\alpha}}{\partial u^a}
$$
together with the tangency conditions
 \begin{eqnarray*}
0&=&Z(\Psi^{\alpha})\vert_{{\mathcal W}_0^{M_C}}=
\left(\frac{\partial \Psi^{\alpha}}{\partial t}+
A^i\frac{\partial \Psi^{\alpha} }{\partial q^i}+
B^a\frac{\partial \Psi^{\alpha} }{\partial u^a}+
C^i\frac{\partial \Psi^{\alpha}}{\partial v^i}\right)\Big\vert_{{\mathcal W}_0^{M_C}}\\
0&=& Z(p+p_iv^i-\costf (t,q,u))\vert_{{\mathcal W}_0^{M_C}} \ .
 \end{eqnarray*}
Therefore the equations of motion are:
\begin{eqnarray*}
\frac{d}{dt}\left(\lambda_{\alpha}(t)\frac{\partial\Psi^{\alpha}}{\partial v^i}(t, q(t),\dot{q}(t),
u(t)) \right)+\frac{\partial \costf }{\partial q^i}(t,q(t),u(t))-
\lambda_{\alpha}(t)\frac{\partial\Psi^{\alpha}}{\partial q^i}(t,q(t),\dot{q}(t), u(t))&=&0\\
\frac{\partial \costf }{\partial u^a}(t,q(t),u(t))-
\lambda_{\alpha}(t)\frac{\partial\Psi^{\alpha}}{\partial u^a}(t, q(t),\dot{q}(t),u(t))&=&0\\
\Psi^{\alpha}(t,q(t),\dot{q}(t), u(t))&=&0
\end{eqnarray*}
Let $\costf_0=\costf-\lambda_{\alpha}\Psi^{\alpha}$ be the
classical extended Lagrangian for constrained systems. Then
these last equations are the usual dynamical equations in
optimal control obtained by applying the Lagrange multipliers method
to the constrained variational
problem, that is, the Euler-Lagrange equations for
$\costf_0$, the extremum necessary condition at interior points, and the constraints.

\newpage

{\bf Remarks}:
\begin{itemize}
\item
In the particular case that $\Psi^j=v^j-{\cal F}^j=0$, the vector field
$Z$ so-obtained is just the image of the vector field obtained in Section \ref{oce}
by the Hamiltonian section (\ref{hamsect}),
as a simple calculation in coordinates shows.
\item
Another obvious but significant remark is that
we can take $\bar\pi^k\colon J^k\pi\to\Real$ (the bundle of $k$-jets of $\pi$)
instead of $\pi\colon E\to\Real$, and hence  $J^k\bar\pi^k$ and
$\Tan ^* J^k\pi$ instead of $J^1\bar\pi^1$ and $\Tan^*E$, respectively.
These changes allows us to address those optimal control problems
where we have $\Phi^{kC}\colon C\to J^k\pi$;
that is, we deal with higher-order equations, and their solutions
must satisfy that $(\gamma(t),j^{k+1}(\pi^k\circ\Phi^{kC}\circ\gamma)(t))\in M$,
where $M$ is a submanifold of $C\times_{J^k\pi}J^{k+1}\pi$.
\end{itemize}

\section{Applications and examples}\label{examples}

\subsection{Optimal Control of Lagrangian systems with controls}

See Appendix \ref{mvfdm} for previous geometric concepts
which are needed in this section. For a complete study of
these systems see \cite{BOV-2002,BLM-2000} and references
therein.

Now we provide a definition of a {\sl controlled-force}, which
allows dependence on time, configuration, velocities and control
inputs. In a global description, one assumes a fiber
bundle structure $\Phi^{1C}: C \longrightarrow J^1\pi$, where $C$
is the bundle of controls, with coordinates $(t,q, v,u)$. Then a
controlled-force is  a smooth map
 ${\bf F}: C\to {\mathcal C}_{\pi}$,
 so that $\pi_{J^1\pi}\circ{\bf F}=\Phi^{1C}$
(see diagram (\ref{yunque})).

In a natural chart, a controlled-force is represented by
\[
{\bf F}(t, q, v, u)= {\bf F}_i(t,q, v, u)(\d q^i-v^i\d t) \
.
\]

A {\sl controlled Lagrangian system} is defined as the pair
$(\Lden, {\bf F})$ which determines an implicit control
system described by the subset $D_C$ of
$C\times_{J^1\pi}J^{2}\pi$:
 \beann D_C&=&\{(c, {\hat p})\in
C\times_{J^1\pi}J^{2}\pi\; |\; (\imath_1^*d_T\Theta_\Lden-
(\pi^2_1)^*\d L )({\hat p})=((\pi^2_1)^*{\bf F})(c)\}
\\ &=&
\{(c, {\hat p})\in C\times_{J^1\pi}J^{2}\pi\; |\; {\cal
E}_\Lden({\hat p})=((\pi^2_1)^*{\bf F})(c)\}
\\ &=&
\{(c, {\hat p})\in C\times_{J^1\pi}J^{2}\pi\; |\; ({\cal
E}_\Lden\circ pr_2-(\pi^2_1)^*{\bf F}\circ pr_1)(c, {\hat
p})=0\}
 \eeann
 where $pr_1$ and $pr_2$ are the natural projections
from $C\times_{J^1\pi}J^{2}\pi$ onto the factors. In fact, $D_C$ is
not necessarily a submanifold of $C\times_{J^1\pi}J^{2}\pi$. There
are a lot of cases where this does happen. In local coordinates
\begin{eqnarray*}
D_C&=&\left\{(t, q, v, w, u)\in C\times_{J^1\pi}J^{2}\pi\;
\Big|\; \frac{\partial^2L}{\partial v^i\partial v^j}(t,
q,v)w^j +
  \frac{\partial^2L}{\partial v^i\partial q^j}(t,q,v)v^j
\right.\\
& & \left. +\frac{\partial^2L}{\partial v^i\partial t}(t,
q, v)-\frac{\partial L }{\partial q^i}(t, q, v) -{\bf
F}_i(t, q, v,u)=0\right\} \ .
 \end{eqnarray*}
A solution to the controlled Lagrangian system $(\Lden,
{\bf F})$ is a map $\gamma\colon\Real\to C$ satisfying
that: \ben
\item[(i)]
$\Phi^{1C}\circ\gamma=j^1(\pi^1\circ\Phi^{1C}\circ\gamma)$.
\item[(ii)]
$(\gamma(t),j^2(\pi^1\circ\Phi^{1C}\circ\gamma)(t))\in
D_C$, for every $t\in\Real$. \een The condition (i) means
that $\Phi^{1C}\circ\gamma$ is holonomic, and (ii) is the
condition (\ref{elecs}) of Appendix \ref{ele}; that is, the
Euler-Lagrange equations for the controlled Lagrangian
system $(\Lden, {\bf F})$.

Now, consider the map $(\hbox{Id}, \Upsilon)\colon
C\times_{J^1\pi}J^2\pi\to C\times_{J^1\pi}J^1\bar\pi^1$, where
$\Upsilon\colon  J^2\pi\to J^1\bar\pi^1$ is defined in
(\ref{upsilon}) (see Appendix \ref{sgs}), and let $M_C=(\hbox{Id},
\Upsilon)(D_C)$. As $(\hbox{Id}, \Upsilon)$ is an injective map, we
can identify $D_C\subset  C\times_{J^1\pi}J^2\pi$ with this subset
$M_C$ of $C\times_{J^1\pi}J^1\bar\pi^1$. Observe that there is a
natural projection from $M_C$ to $J^1\pi$.

If $\costf \colon M_C\to \R$ is  a cost function, we may consider
the implicit optimal control system determined by the pair
$(\Lag, M_C)$, where $\Lag=\costf\d t$,
and apply the method developed in Section \ref{iocp}.

Let $\overline{\mathcal{W}}^{M_C}=M_C\times_{J^1\pi} T^* J^1\pi$,
and
$\overline{\mathcal{W}}^C=C\times_{J^1\pi}J^1\bar\pi^1\times_{J^1\pi}
T^* J^1\pi$. The natural projection from $\overline{\mathcal{W}}^C$
to $T^* J^1\pi$ allows us to pull-back the canonical $2$-form
$\Omega_{J^1\pi}$ to a presymplectic form
$\Omega_{\overline{\mathcal{W}}^C}\in\df^2(\overline{\mathcal{W}}^C)$.
Furthermore, in $J^1\bar\pi^1\times_{J^1\pi} T^* J^1\pi$ there is
the natural coupling form $\bar{\hat{\cal C}}$ (see Definition
\ref{coupling}). We denote by $\bar{\cal C}$ its pull-back to
$\overline{\mathcal{W}}^C$. We denote by $\Lag$ and $\costf$ the
pull-back of $\Lag$ and $\costf$ from $M_C$ to
$\overline{\mathcal{W}}^C$, for the sake of simplicity.

Then, let $\bar H_{\mathcal{W}^C}\colon \overline{\mathcal{W}}^C\to
\R$ be the unique function such that $\bar{\mathcal C}-\Lag=\bar
H_{\mathcal{W}^C}\d t$, whose local expression is $\bar
H_{\mathcal{W}^C}=p+p_i\bar v^i+\bar p_iw^i-\costf$, and consider
the submanifold $\overline{\cal W}_0=\{\tilde
q\in\overline{\mathcal{W}}^C \ \vert\ \bar H_{\mathcal{W}^C}(\tilde
q)~=~0\}$. The pull-back of $\bar H_{\mathcal{W}^C}$ to
$\overline{\mathcal{W}}^{M_C}$ is the Pontryagin Hamiltonian,
denoted by $\bar H_{\mathcal{W}^{M_C}}$.

Finally, the dynamics is in the submanifold
$\overline{\mathcal{W}}^{M_C}_0=\overline{\mathcal{W}}^{M_C}\cap\overline{\cal
W}_0$ of $\overline{\mathcal{W}}^C$, where $\jmath_1^{M_C}$ is the
natural embedding. $\overline{\mathcal{W}}^{M_C}_0$ is endowed with
the presymplectic form
$\Omega_{\overline{\mathcal{W}}^{M_C}_0}=(\jmath_1^{M_C})^*\Omega_{\overline{\mathcal{W}}^C}$.
Therefore, the motion is determined by a vector field
$Z\in\vf(\overline{\mathcal{W}}^{M_C}_0)$ satisfying the equations
$$
\inn(Z)\Omega_{\overline{\mathcal{W}}_0^{M_C}}=0\quad ,\quad
\inn(Z)\d t=1 \ .
$$

A local chart in $\overline{\mathcal{W}}^C$ is
$(t,q^i, v^i,\bar v^i,w^i,u^a, p, p_i,\bar{p}_i)$,
where $(\bar v^i,w^i)$ and $(p, p_i,\bar{p}_i)$ are the natural fiber coordinates in
 $J^1\bar\pi^1$ and $\Tan^*J^1\pi$, respectively.
The manifold $\overline{\mathcal{W}}^{M_C}$ is given locally by the $2n$ constraints:
\begin{eqnarray*}
\varphi_i(t,q^i, v^i,\bar v^i,w^i,u^a, p, p_i,\bar{p}_i)&=&
w^j\frac{\partial^2L}{\partial v^i\partial v^j}(t, q,v) +
\bar{v}^j\frac{\partial^2L}{\partial v^i\partial q^j}(t, q,v)+
\frac{\partial^2L}{\partial v^i\partial t}(t, q, v)
\\ &&
-\frac{\partial L}{\partial q^i}(t, q,v)-{\bf F}_i(t,q, v,
u)=0
\\
\bar{\varphi}^i(t,q^i, v^i,\bar v^i,w^i,u^a, p, p_i,\bar{p}_i)&=&v^i-\bar{v}^i=0 \ ,
\end{eqnarray*}
and $\overline{\mathcal{W}}_0$ is given by
$$
\phi(t,q^i, v^i,\bar v^i,w^i,u^a, p, p_i,\bar{p}_i)= \bar
H_{\mathcal{W}^C}(t,q^i, v^i,\bar v^i,w^i,u^a, p,
p_i,\bar{p}_i)= p+p_i\bar{v}^i+\bar{p}_i w^i-\costf (t,q, v,u)=0 \ ,
$$
and
\[
\Omega_{\overline{\mathcal{W}}^{M_C}_0}=
\d q^i\wedge\d p_i +\d v^i\wedge\d\bar{p}_i+\d t\wedge\d (\costf-p_i\bar v^i-\bar p_iw^i) \ .
\]

Following Proposition \ref{2cons}, we look for a vector field
 $Z\in\vf(\overline{\mathcal{W}}^C)$ such that,
for every ${\bf w}\in\overline{\mathcal{W}}^{M_C}_0$:
$$
{\rm (i)} \quad Z_{\bf w}\in\Tan_{\bf
w}\overline{\mathcal{W}}^{M_C}_0 \quad ,\quad {\rm (ii)}\quad
\inn(Z_{\bf w})\Omega_{\overline{\mathcal{W}}^C}\in(\Tan_{\bf
w}\overline{\mathcal{W}}^{M_C}_0)^0 \ ,
$$
or, equivalently
\ben
\item[(i)]
$(\jmath_1^{M_C})^*(Z(\varphi_i))=0$,\quad
$(\jmath_1^{M_C})^*(Z(\bar\varphi^i))=0$,\quad
$(\jmath_1^{M_C})^*(Z(\phi))=0$.
\item[(ii)]
$(\jmath_1^{M_C})^*(\inn(Z)\Omega_{\overline{\mathcal{W}}^C})=0$.
\een
 Remember that the constraints are $\varphi_i=0$,\quad
$\bar\varphi^i=0$,\quad $\phi=0$.

If $Z$ is given locally by
\[
Z=\frac{\partial}{\partial t}+A^i\frac{\partial}{\partial q^i}+
{\mathcal A}^i\frac{\partial}{\partial v^i}+
\bar{A}^i\frac{\partial}{\partial\bar{v}^i}+
\bar{\mathcal{A}}^i\frac{\partial}{\partial w^i}+
B^a\frac{\partial}{\partial u^a}+D\frac{\partial}{\partial p}+
 C_i\frac{\partial}{\partial p_i}+\bar{C}_i\frac{\partial}{\partial \bar{p}_i}\; ,
\]
then
 $A^i, {\mathcal A}^i, \bar{A}^i,\bar{\mathcal{A}}^i, B^a,D, C_i, \bar{C}_i$
 are unknown functions in $\overline{\mathcal W}^C$, such that
\[
\inn(Z)\Omega_{\overline{\mathcal{W}}^C}=\lambda^i \d\varphi_i +
\bar{\lambda}_i \d\bar{\varphi}^i+\lambda \d(p+p_i
\bar{v}^i+\bar{p}_iw^i-\costf (t,q, v, u))
\]
and $Z(\varphi_i)=0$, $Z(\bar{\varphi}^i)=0$ and
 $Z(p+p_i\bar{v}^i+\bar{p}_i w^i-\costf (t,q, v,u))=0$.
  From these equations we obtain
\bea
\lambda=1 \ , &
A^i= \bar{v}^i  \ ,&
\mathcal{A}^i= w^i \nonumber
\\
C_i=\frac{\partial\costf }{\partial q^i}-\lambda^j\frac{\partial\varphi_j}{\partial q^i}\ , &
\bar{C}_i=
\ds\frac{\partial\costf}{\partial v^i}-\lambda^j\frac{\partial \varphi_j}{\partial v^i}-\bar{\lambda}_i\ , &
D=\frac{\partial\costf}{\partial t}-\lambda^j\frac{\partial\varphi_j}{\partial t} \nonumber
\\
0=\frac{\partial\costf}{\partial
u^{a}}+\lambda^i\frac{\partial{\bf F}_i}{\partial u^a}\ , &
p_i=\ds\bar{\lambda}_i-\lambda^j\frac{\partial^2 L
}{\partial v^j\partial q^i}\ , &
\bar{p}_i=-\lambda^j\frac{\partial^2 L }{\partial
v^i\partial v^j} \label{p-i} \eea and the tangency
conditions \bea Z(\varphi_i)&=&\frac{\partial
\varphi_i}{\partial t}+ \bar{v}^j\frac{\partial
\varphi_i}{\partial q^j}+w^j\frac{\partial
\varphi_i}{\partial v^j}+ \bar A^j\frac{\partial^2 L
}{\partial v^i\partial q^j}- B^a\frac{\partial {\bf
F}_i}{\partial u^a}+
\bar{\mathcal{A}}^j\frac{\partial^2 L }{\partial v^i\partial v^j}=0\label{cde0}\\
Z(\bar{\varphi}^i)&=&w^i-\bar{A}^i=0 \nonumber\\
Z(\phi)&=&Z(p+p_i \bar{v}^i+\bar{p}_i w^i-\costf (t,q,v, u))=0 \nonumber
\eea
where the third condition is satisfied identically using the previous equations.

Assuming that the Lagrangian $L$ is regular, that is,
$\det(W_{ij})=det\left(\frac{\partial^2 L}{\partial v^i\partial v^j}\right)\not= 0$,
 then from equations for $p_i$ and $\bar p_i$ in (\ref{p-i})
we obtain explicit values of the Lagrange multipliers $\lambda^i$ and $\bar\lambda_i$.
Therefore, the remaining equations (\ref{p-i}) are
now rewritten as the new set of constraints
\beq
\label{nose}
 \psi^a(t,q,v,u,\bar{p})=
\frac{\partial \costf }{\partial
u^{a}}-W^{ij}\bar{p}_i\frac{\partial {\bf F}_j}{\partial
u^a}=0 \ , \eeq which corresponds to $\ds \derpar{\bar
H_{\mathcal{W}^{M^C}}}{u^a}=0$.

The new compatibility condition is
\begin{equation}
\label{cde} Z(\psi^a)=\frac{\partial \psi^a}{\partial
t}+\bar{v}^j\frac{\partial \psi^a}{\partial q^j}+
w^j\frac{\partial\psi^a}{\partial
v^j}+B^b\frac{\partial\psi^a}{\partial u^b}+ \bar{C}_i\frac{\partial
\psi^a }{\partial\bar{p}_i}=0 \ .
\end{equation}
Furthermore we assume that
\[
\det \left(\frac{\partial \psi^a}{\partial u^b}\right)\not= 0 \ ,
\]
then, from Equations (\ref{cde0}) and (\ref{cde})
we obtain the remaining components $\bar{\mathcal
A}^i$ and $B^a$, and we determine completely the vector field $Z$.

The equations of motion for a curve are
determined by the system of implicit-differential
equations:
\begin{eqnarray}
\dot{p}_i(t)&=&\frac{\partial \costf }{\partial q^i}(t,q(t), \dot{q}(t),
u(t))-\lambda^j(t,q(t), \dot{q}(t),
\bar{p}(t))\frac{\partial\varphi_j}{\partial q^i}(t,q(t), \dot{q}(t),\ddot{q}(t),u(t))
\label{q1q} \nonumber\\
\dot{\bar{p}}_i(t)&=&\frac{\partial \costf }{\partial v^i}(t,q(t), \dot{q}(t),u(t))-p_i(t)
\nonumber\\
&& -\lambda^j(t,q(t), \dot{q}(t),\bar{p}(t))
\left[\frac{\partial \varphi_j}{\partial v^i}(t,q(t), \dot{q}(t), \ddot{q}(t),u(t))+
\frac{\partial^2L}{\partial v^j\partial q^i}(t,q(t), \dot{q}(t))\right]\label{q2q}\\
0&=&\frac{d}{dt}\left(\frac{\partial L}{\partial v^i}(t,q(t), \dot{q}(t))\right)-
\frac{\partial L}{\partial q^i}(t,q(t), \dot{q}(t))-{\bf F}_i(t,q(t), \dot{q}(t), u(t))\label{q3q}\\
0&=&\frac{\partial \costf }{\partial u^{a}}(t,q(t),
\dot{q}(t),u(t))-W^{ij}(t,q(t),\dot{q}(t))\bar{p}_i(t)\frac{\partial{\bf
F}_j}{\partial u^a}(t,q(t), \dot{q}(t), u(t)) \ .
\label{q4q}
\end{eqnarray}
Equation (\ref{q4q}) is the explicit expression of (\ref{nose}).

In \cite{AF} the authors study optimal control of
Lagrangian systems with controls in a more restrictive
situation using higher-order dynamics, obtaining that the
states are determined by a set of fourth-order differential
equations. First it is necessary to assume that the system
is {\sl fully actuated}, that is $m=n$, and
$\hbox{rank}\left(
\Xi_{ij}\right)=\hbox{rank}\left(\frac{\partial {\bf
F}_i}{\partial u^j}\right)=n$. Moreover, in the sequel we
assume that the system is affine on controls, that is,
\[
{\bf F}_i(t, q, \dot{q}, u)=A_i(t, q, \dot{q})+ A_{ij}(t,
q, \dot{q})u^j \ .
\]
Therefore, $\Xi_{ij}=A_{ij}$.

Then from the constraint equations (\ref{q3q}) and (\ref{q4q}),
 applying the Implicit Function Theorem,  we deduce that
\begin{eqnarray*}
u^i(t)&=&u^i(t,q(t),
\dot{q}(t),
\ddot{q}(t))=A^{ij}\left[
\frac{d}{dt}\left(\frac{\partial L}{\partial v^j}(t,q(t),
\dot{q}(t))\right)-\frac{\partial L}{\partial q^j}(t,q(t),
\dot{q}(t))-A_j(t,q(t), \dot{q}(t))    \right]\\
\bar{p}_i(t)&=&{\mathcal H}^j_i(t,q(t), \dot{q}(t))\frac{\partial
\costf }{\partial u^{j}}(t,q(t),\dot{q}(t), u(t,q(t),\dot{q}(t),
\ddot{q}(t)))
\end{eqnarray*}
where $({\mathcal H}^j_i)$ are the components of
the inverse matrix of the regular matrix
$(W^{ik}A_{kj})$.

Taking the derivative  with respect to time of Equation
(\ref{q2q}), and substituting the value of $\dot{p}_i(t)$ using
Equation (\ref{q1q}) we obtain a fourth-order differential
equation depending on the states. After some computations we
deduce that
\[
{\mathcal H}_i^j(t,q(t), \dot{q}(t))\frac{\partial^2\costf
}{\partial u^{j}\partial{u}^k}(t,q(t), \dot{q}(t),
\ddot{q}(t))\frac{d^4 q^k}{dt^4}(t)=G_i(t,q(t), \dot{q}(t),
\ddot{q}(t), \dddot{q}(t)) \ .
\]
Finally, under the assumption that the matrix
$\left(\frac{\partial^2 \costf }{\partial
u^{j}\partial{u}^k}\right)$ is invertible, we
obtain a explicit fourth-order system of
differential equations:
\[
\frac{d^4 q^i}{dt^4}(t)=\bar{G}^i(t,q(t), \dot{q}(t), \ddot{q}(t),
\dddot{q}(t)) \ .
\]

\subsection{Optimal Control  problems for descriptor systems}

See \cite{muller} for the origin and interest of this example.
The study of these kinds of systems was suggested to us by
Professor. A.D. Lewis (Queen's University of Canada).

Consider the problem of minimizing the functional
\[
{\mathcal J}=\frac{1}{2}\int_0^{+\infty}\left[ a_i(q^i)^2+ru^2\right]\, \d t,
\]
$1\leq i\leq 3$, with control equations
$$
\dot{q}^2=q^1+b_1 u\quad ,\quad
\dot{q}^3=q^2+b_2 u\quad ,\quad
0= q^3+b_3 u
$$
with parameters $a_i, b_i\geq 0$ and $r>0$.

As in the previous section, the geometric
framework developed in Section  \ref{ocpg} is
also valid for this class of systems.
Let $E=\R\times \R^3$ with coordinates $(t, q^i)$, and
$C=\R\times\R^3\times\R$ with coordinates $(t,q^i, u)$.
The submanifold $M_C\subset C\times_E J^1\pi$ is given by
\[
M_C=\{ (t, q^1, q^2, q^3, v^1, v^2, v^3, u)\; |\; v^2=q^1+b_1 u\; ,
v^3=q^2+b_2 u\; , 0= q^3+b_3 u \} \, .
\]
 The cost function is
\[
\begin{array}{rrcl}
\costf :& C &\longrightarrow& \R\\
  &(t, q^1, q^2, q^3, u)&\longmapsto& \displaystyle{\frac{1}{2}\left[ a_1
  (q^1)^2+  a_2  (q^2)^2 +  a_3  (q^3)^2+ru^2\right]}
\end{array}
\]
We analyze the dynamics of the implicit optimal control system
determined by the pair $(\Lag , M_C)$.

Let $\mathcal{W}^{M_C}=M_C\times_E T^*E$ and
$\mathcal{W}^C=C\times_E J^1\pi \times_E T^*E$ with coupling form
${\cal C}$ inherited from the natural coupling form in $J^1\pi
\times T^*E$. Let $H_{\mathcal{W}^C}\colon \mathcal{W}^C \to \R$ be
the unique function such that $\mathcal{C}-\Lag=H_{\mathcal{W}^C}\d
t$, and consider the submanifold ${\cal W}_0=\{\tilde q\in
\mathcal{W}^C \ \vert H_{\mathcal{W}^C}(\tilde q)=0\}$. Finally, the
dynamics is in the submanifold
$\mathcal{W}^{M_C}_0=\mathcal{W}^{M_C}\cap{\cal W}_0$ of
$\mathcal{W}^C$. Locally,
\beann
 {\mathcal W}_0^{M_C} &=&\{
(t,q^1,q^2,q^3,v^1,v^2,v^3,u,p,p_1,p_2,p_3)\ |\ v^2=q^1+b_1
u\; , v^3=q^2+b_2 u\; ,
\\  & & \; q^3+b_3 u= 0\; , p+p_1v^1+p_2v^2+p_3v^3-\costf=0 \}\, .
 \eeann

Therefore, the motion is determined by a vector field
$Z\in\vf(\mathcal{W}^{M_C}_0)$ satisfying the Equations
(\ref{vhs1}), which according to Proposition
\ref{2cons} is equivalent to finding a vector field $Z\in \vf
(\mathcal{W}^C)$ (if it exists):
\[
Z=\frac{\partial}{\partial t}+ A^1\frac{\partial}{\partial
q^1}+A^2 \frac{\partial}{\partial q^2}+A^3\frac{\partial}{\partial
q^3}+ C^1\frac{\partial}{\partial v^1}+C^2\frac{\partial}{\partial
v^2}+C^3\frac{\partial}{\partial v^3}+B\frac{\partial}{\partial
u}+
 D_1\frac{\partial}{\partial p_1}+ D_2\frac{\partial}{\partial p_2}+
D_3\frac{\partial}{\partial p_3}+ E\frac{\partial}{\partial p}
\] such that
\beann \inn(Z)\Omega_{\mathcal{W}^C}=\lambda_1
d(q^1+b_1u-v^2)+\lambda_2
d(q^2+b_2u-v^3)+\lambda_3 d(q^3+b_3u)+\lambda d H_{\mathcal{W}^C}\; , \\
Z(q^1+b_1u-v^2)= 0 \; , \quad Z(q^2+b_2u-v^3)=0 \; ,\quad Z(q^3+b_3
u)=0\; , \quad Z(H_{\mathcal{W}^C})=0 \eeann where
$\Omega_{\mathcal{W}^C}\in \df^2(\mathcal{W}^C)$ is the 2-form with
local expression
\[
\Omega_{\mathcal{W}^C}=\d q^1\wedge \d p_1+ \d q^2\wedge \d p_2 + \d
q^3\wedge \d p_3+\d t \wedge \d p \; .\]

After some straightforward computations, we obtain that
\beann
A^1=v^1\quad , & A^2=q^1+b_1u  \quad & , \quad  A^3=q^2+b_2 u \\
\lambda=1 \quad , & E=0 \quad & , \quad
0=ru-b_1p_2-b_2p_3-b_3\lambda_3
\\
C^2=v^1+b_1B \quad , & C^3=A^2+b_2 B \quad & , \quad 0=A^3+b_3B \\
p_1=0 \quad , & p_2=\lambda_1 \quad  & , \quad p_3=\lambda_2 \\
  D_1=a_1q_1-p_2\quad , & D_2=a_2q_2-p_3 \quad & , \quad
  D_3=a_3q_3-\lambda_3 \, .
\eeann We deduce that
\[
\lambda_3= \frac{1}{b_3} (ru-b_1p_2-b_2p_3)\; ,\quad
B=-\frac{1}{b_3}(q^2+b_2u)\; .
\]
Therefore, the new constraint submanifold
${\mathcal W}_1^{M_C}\hookrightarrow
{\mathcal W}_0^{M_C}$ is
\[
{\mathcal W}_1^{M_C}=\{(t,q^1,q^2,v^1,u,p_1,p_2,p_3)\; |\; p_1=0\}
\, .
\]
Consistency of the dynamics implies that
\[
0=Z(p_1)=D_1=a_1q_1-p_2 \, .
\]
Thus,
\[ {\mathcal W}_2^{M_C}=\{(t,q^1,q^2,v^1,u,p_2,p_3)\; |\; a_1q_1-p_2=0\}
\]
and once again we impose the tangency to the new constraints:
\[
0=Z(a_1q_1-p_2)=a_1v^1-a_2q_2+p_3
\]
which implies that
\[ {\mathcal W}_3^{M_C}=\{(t,q^1,q^2,v^1,u,p_3)\; |\;
a_1v^1-a_2q^2+p_3=0 \} \, .
\]
{}From  the compatibility condition
\[ 0=Z(a_1v^1-a_2q^2+p_3)
\]
and the constraints we determine the remaining component $C^1$ of
$Z$:
\[
C^1=
\frac{1}{a_1b_3}\left[(a_2b_3-a_1b_1)q^1-b_2a_2q^2+(a_2b_1b_3+a_3b_3^2+r)u+b_2a_1v^1\right]
\, .
\]
Therefore the equations of motion of the optimal control problem
are:
\begin{eqnarray}
\ddot{q}^1(t)&=&\frac{1}{a_1b_3}\left[(a_2b_3-a_1b_1)q^1(t)-a_2b_2q^2(t)+(a_2b_1b_3+a_3b_3^2+r)u(t)+
a_1b_2\dot{q}^1(t)\right]\label{muller}\\
\dot{q}^2(t)&=&q^1(t)+b_1 u(t)\nonumber\\
0&=&q^2(t)+b_2 u(t)-b_3\dot{u}(t) \, .\nonumber
\end{eqnarray}
{}From (\ref{muller}) we deduce that
\[
u(t)=\frac{1}{a_2b_1b_3+a_3b_3^2+r}\left[(a_1b_1-a_2b_3)q^1(t)+a_2b_2q^2(t)-a_1b_2\dot{q}^1(t)+
a_1b_3\ddot{q}^1(t)\right] \, .
\]
This is the result obtained in M\"{u}ller
\cite{muller}, where the optimal feedback control
depends on the state variables and also on their
derivatives (non-casuality).

Choosing local coordinates $(t, q^1, q^2, v^1,u)$ on ${\mathcal W}_3^{M_C}$,
 if $\jmath_3:{\mathcal W}_3^{M_C}\mapsto {\mathcal W}^C$
 is the canonical embedding, then
$\Omega_{{\mathcal W}_3^{M_C}}=\jmath_3^*\Omega_{\mathcal{W}^C}$ is
locally written as
\[
\Omega_{{\mathcal W}_3^{M_C}}=-a_1 \d q^1\wedge \d q^2+a_2b_3 \d
q^2\wedge \d u-a_1b_3 \d v^1\wedge \d u+\d t\wedge  \d \jmath_3^*p
\, ,
\] where $\jmath_3^*p: {\mathcal W}_3^{M_C}\to \R$ is
the function
\[
\jmath_3^*p=-\frac{1}{2}a_1(q^1)^2-\frac{1}{2}a_2(q^2)^2+\frac{1}{2}(r+a_3b^2_3)u^2-a_1b_1
q^1u-a_2b_2 q^2 u+a_1b_2v^1u+a_1q^2v^1 \, .
\]
Obviously, $(\Omega_{{\mathcal W}_3^{M_C}}, \d t)$ is a
cosymplectic structure on ${\mathcal W}_3^{M_C}$ (see
Proposition \ref{cosympl}), and there exists a  unique
vector field $\bar{Z}\in \vf({\mathcal W}_3^{M_C})$
satisfying
\[
\inn({\bar{Z}}) \Omega_{{\mathcal W}_3^{M_C}}=0,\quad
\inn({\bar{Z}})\d t=1 \, .
\]

\section{Conclusions and outlook}

In this paper we have elucidated the geometrical structure of
optimal control problems  using  a variation of the Skinner-Rusk
formalism for mechanical systems. The geometric framework allows us
to find the dynamical equations of the problem (equivalent to the
Pontryagin Maximum Principle for  smooth enough problems
without boundaries on the space of controls), and
to describe the submanifold (if it exists) where the solutions of
the problem are consistently defined. The method admits a nice
extension for studying the dynamics of implicit
optimal control problems with a wide range of applicability.

One line of future research appears when we combine our geometric
method for optimal control problems, and the study of the
(approximate) solutions to optimal control problems involving
partial differential equations when we  discretize the space
domain and consider the resultant set of ordinary differential
equations (see, for instance, \cite{Br} and references therein and
\cite{LMM-2007}, for a geometrical description). This resultant
system is an optimal control problem, where the state equations
are, presumably, a very large set of coupled ordinary
differential equations. Typically, difficulties other than computational ones
appear because the system is
differential-algebraic, and therefore the optimal
control problem is a usual one for a descriptor system.

Moreover, in this paper we have confined ourselves to the
geometrical aspects of time-dependent optimal control problems. Of
course,  the techniques are suitable for studying the formalism for
optimal control problems for partial differential equations in general.

\appendix
\section{Appendix}
\protect\label{mvfdm}

\subsection{Tulczyjew's operators}

Given a differentiable manifold $Q$ and its tangent bundle
$\tau_Q\colon\Tan Q\to Q$, we consider
the following operators, introduced by Tulczyjew \cite{Tulczy1}:
first we have
$\inn_T\colon\df^k(Q)\longrightarrow\df^{k-1}(\Tan Q)$, which is defined as follows:
for every $({\rm p},v)\in\Tan Q$, $\alpha\in\df^k(Q)$, and $\moment{X}{1}{k-1}\in\vf(\Tan Q)$,
$$
(\inn_T\alpha)(({\rm p},v);\moment{X}{1}{k-1})=
\alpha({\rm p};v,\Tan_{({\rm p},v)}\tau_Q((X_1)_{({\rm p},v)}),\ldots,
\Tan_{({\rm p},v)}\tau_Q((X_{k-1})_{({\rm p},v)})) \ .
$$
Then, the so-called {\sl total derivative} is a map
$\d_T\colon\df^k(Q)\to\df^k(\Tan Q)$ defined by
$$
d_T=\d\circ\inn_T+\inn_T\circ\,\d \ .
$$
For the case $k=1$, using natural coordinates in $\Tan Q$, the local expression is
$$
d_T\alpha\equiv \d_T(A_j\d q^j)=A_j\d v^j+v^i\derpar{A_j}{q^i}\d q^j \ .
$$

\subsection{Some geometrical structures}
\protect\label{sgs}

Recall that, associated with every jet bundle $J^1\pi$, we have
the {\sl contact system}, which is a subbundle
 ${\mathcal C}_{\pi}$ of $T^*J^1\pi$ whose fibres
at every $j^1\phi(t)\in J^1\pi$ are defined as
\[
{\mathcal C}_{\pi}(j^1\phi(t)) =
 \{ \alpha\in \Tan_{j^1\phi(t)}^*(J^1\pi)\; |\;
\alpha=(\Tan_{j^1\phi(t)}\pi^1-\Tan_{j^1\phi(t)}(\phi\circ\bar\pi^1))^*\beta,\
\beta\in{\rm V}^*_{\phi(t)}\pi\}) \ .
\]
One may readily see that a local basis for the sections of this bundle is given by
$\{\d q^i-v^i\d t\}$.

Now, denote by $J^2\pi$ the bundle of 2-jets of $\pi$.
This jet bundle is equipped with
natural coordinates $(t, q^i, v^i, w^i)$ and canonical projections
\[
\pi^2_1\colon J^2\pi\to J^1\pi,\quad \pi^2\colon J^2\pi\to E\quad
, \quad \bar{\pi}^2\colon J^2\pi\to \R \ .
\]
Considering the bundle $J^1\bar{\pi}^1$,
we introduce the canonical injection
 $\Upsilon\colon J^2\pi \to J^1\bar{\pi}^1$
given by
\beq
\label{upsilon}
\Upsilon (j^2\phi(t))=(j^1(j^1\phi))(t)\, .
\eeq
Taking  coordinates $(t, q^i, v^i;  \bar{v}^i, w^i)$ in
$J^1\bar{\pi}^1$ then
$\Upsilon(t, q^i, v^i, w^i)=(t, q^i, v^i;  v^i, w^i)$.

Thus, we have the following diagram
\beq
\bfig\xymatrix{
\Tan J^1\pi=\Tan\Real\times\Tan(\Tan Q)
\ar[ddrr]_{\txt{\small{$\tau_{J^1\pi}$}}}&&
T^*(J^2\pi)
\ar[d]_{\txt{\small{$\pi_{J^2\pi}$}}}
&& J^1\bar\pi^1=\Real\times\Tan(\Tan Q)
\ar[ddll]^{\txt{\small{$(\bar\pi^1)^1$}}} \\
 && J^2\pi=\Real\times\Tan^2Q
\ar[d]_{\txt{\small{$\pi_1^2$}}}
\ar[ull]_{\txt{\small{$\imath_ 1$}}}
\ar[urr]^{\txt{\small{$\Upsilon$}}} &&  \\
 && J^1\pi=\Real\times\Tan Q
 \ar[lldd]_{\txt{\small{$\pi^1$}}}
\ar[rrdd]^{\txt{\small{$\bar\pi^1$}}}
&& \\
 & \qquad {\cal C}_\pi  \ar@^{(->}[r]& \Tan^*J^1\pi
\ar[u]^{\txt{\small{$\pi_{J^1\pi}$}}} & &
\\
 \Real\times Q \ar[rrrr]^{\txt{\small{$\pi$}}} & & & & \Real
}
\efig
\label{yunque}
\eeq
where the inclusion $\imath_1$ is locally given by
$\imath_1(t,q,v,w)=(t,1,q,v,v,w)$.

Observe that $(\pi^2_1)^*\Tan^*J^1\pi$ can be identified with
a subbundle of $\Tan^*J^2\pi$ by means of the natural injection
$\hat \i\colon (\pi^2_1)^*\Tan^*J^1\pi\to\Tan^*J^2\pi$,
defined as follows: for every
$\hat p\in J^2\pi$, $\alpha\in\Tan^*_{\pi^2_1(\hat p)}J^1\pi$,
and ${\rm a}\in\Tan_{\hat p}J^2\pi$,
$$
(\hat \i(\hat p,\alpha))({\rm a})=\alpha(\Tan_{\hat p}\pi^2_1({\rm a}))\ .
$$
In the same way, we can identify
$(\pi^2_1)^*{\mathcal C}_{\pi}$ as a subbundle of
$(\pi^2_1)^*\Tan^*J^1\pi$ by means of $\hat \i$.

Local bases for the set of sections of the bundles
$\Tan^*J^2\pi\to J^2\pi$, $(\pi^2_1)^*\Tan^*J^1\pi\to J^2\pi$,
and $(\pi^2_1)^*{\mathcal C}_{\pi}\to J^2\pi$
are $(\d t,\d q^i,\d v^i,\d w^i)$, $(\d t,\d q^i,\d v^i)$, and $(\d q^i-v^i\d t)$,
respectively.

Incidentally,
$Sec\,(J^2\pi,(\pi_1^2)^*\Tan^*J^1\pi)=
\Cinfty(J^2\pi)\otimes_{\Cinfty(J^1\pi)}(\pi_1^2)^*\df^1(J^1\pi)$,
which are the $\pi_1^2$-semibasic $1$-forms in $J^2\pi$.

\subsection{Euler-Lagrange equations}
\protect\label{ele}

Let $\Lden\in \df^1(J^1\pi)$ be a  Lagrangian density and
its associated Lagrangian function $L\in\Cinfty(J^1\pi)$.
Observe that
$$
d_T\Theta_\Lden\in\df^1(\Tan J^1\pi) \ , \
\imath_1^*d_T\Theta_\Lden\in\df^1(J^2\pi) \ , \
(\pi^2_1)^*\d L\in\df^1(J^2\pi) \ .
$$
Then, a simple calculation in coordinates shows that
$\imath_1^*d_T\Theta_\Lden-(\pi^2_1)^*\d L$
is a section of the bundle projection
$\hat \i((\pi^2_1)^*{\mathcal C}_{\pi})\to J^2\pi$.

The Euler-Lagrange equations for this Lagrangian are a  system of second order
differential equations on $Q$; that is, in implicit form, a
submanifold $D$ of $J^{2}\pi$ determined by:
$$
D=\{\hat p\in J^{2}\pi\; |\; (\imath_1^*d_T\Theta_\Lden-(\pi^2_1)^*\d L )(\hat p)=0\}=
\{\hat p\in J^{2}\pi\; |\; {\cal E}_\Lden(\hat p)=0\}={\cal E}_\Lden^{-1}(0)\ ,
$$
where ${\cal E}_\Lden=\imath_1^*d_T\Theta_\Lden-(\pi^2_1)^*\d L$.
Then, a section $\phi\colon\Real\to\Real\times Q$ is a solution to the
Lagrangian system  if, and only if,
${\rm Im}\, j^2\phi\subset{\cal E}_\Lden^{-1}(0)$.
 In fact, working in local coordinates, such as
\beann
d_T\Theta_\Lden &=&
\derpar{L}{v^k}\d v^k-\left(\derpar{L}{v^j}v^j-L\right)\d \dot t
+\left(\dot t\frac{\partial^2L}{\partial t\partial v^k}+
v^i\frac{\partial^2L}{\partial q^i\partial v^k}+
w^i\frac{\partial^2L}{\partial v^i\partial v^k}\right)\d q^k \\ & &
-\left[\dot t\left( v^j\dot t\frac{\partial^2L}{\partial t\partial v^j}-\derpar{L}{t}\right)+
 v^i\left(v^j\frac{\partial^2L}{\partial q^i\partial v^j}-\derpar{L}{q^i}\right)+
w^i\left(\derpar{L}{v^i}+v^j\frac{\partial^2L}{\partial v^i\partial v^j}-\derpar{L}{v^i}\right)\right]\d t
\\
\imath_1^*d_T\Theta_\Lden &=&
\derpar{L}{v^k}\d v^k
+\left(\frac{\partial^2L}{\partial t\partial v^k}+
v^i\frac{\partial^2L}{\partial q^i\partial v^k}+
w^i\frac{\partial^2L}{\partial v^i\partial v^k}\right)\d q^k \\ & &
-\left[v^j\frac{\partial^2L}{\partial t\partial v^j}-\derpar{L}{t}+
 v^i\left(v^j\frac{\partial^2L}{\partial q^i\partial v^j}-\derpar{L}{q^i}\right)+
w^iv^j\frac{\partial^2L}{\partial v^i\partial v^j}\right]\d t
\\
(\pi^2_1)^*\d L &=&
\derpar{L}{t}\d t+\derpar{L}{q^k}\d q^k+\derpar{L}{v^k}\d v^k \ ,
\eeann
we obtain
\beann
\imath_1^*d_T\Theta_\Lden-(\pi^2_1)^*\d L &=&
\left(\frac{\partial^2  L}{\partial v^i\partial v^k}w^i+\frac{\partial^2  L}{\partial q^i\partial v^k}v^i+
\frac{\partial^2  L}{\partial t\partial v^k}-\frac{\partial L}{\partial q^k}\right)
(\d q^k-v^k\d t) \\ &=&
\left[\frac{d}{d t}\left(\frac{\partial L}{\partial v^k}\right)-\frac{\partial  L}{\partial q^k}\right]
(\d q^k-v^k\d t) \ .
\eeann

Now, suppose that
there are external forces operating on the Lagrangian system $(J^1\pi,\Lden)$.
A force depending on velocities is a section $F\colon J^1\pi\to {\cal C}_\pi$.
As above, the corresponding Euler-Lagrange equations are a system of second order
differential equations on $Q$, given in implicit form by the
submanifold $D_F$ of $J^{2}\pi$ determined by:
$$
D_F=\{\hat p\in J^{2}\pi\; |\; (\imath_1^*d_T\Theta_\Lden-(\pi^2_1)^*\d L )(\hat p)=
(F\circ\pi^2_1)(\hat p)\}=
\{\hat p\in J^{2}\pi\; |\; {\cal E}_\Lden(\hat p)=(F\circ\pi^2_1)(\hat p)\}\ .
$$
A section $\phi\colon\Real\to\Real\times Q$ is a solution to the
Lagrangian system  if, and only if,
\beq
{\cal E}_\Lden( j^2\phi)=(\pi^2_1)^*[(F\circ\pi^2_1)( j^2\phi)]=
(\pi^2_1)^*F( j^1\phi)\ .
\label{elecs}
\eeq
In natural coordinates we have
$$
\left[\frac{d}{d t}\left(\frac{\partial L}{\partial v^k}\right)-\frac{\partial  L}{\partial q^k}\right]
(\d q^k-v^k\d t)=
F_j(\d q^j-v^j\d t) \ .
$$

\subsection*{Acknowledgments}

We acknowledge the financial support of \emph{Ministerio de
Educaci\'on y Ciencia}, Projects MTM2005-04947,
MTM2004-7832, and  S-0505/ESP/0158 of the CAM. One of us (MBL)
also acknowledges the financial support of the FPU grant
AP20040096. We thank Mr. Jeff Palmer for his assistance in
preparing the English version of the manuscript.


\begin{thebibliography}{99}

\begin{small}
\bibitem{AF}
{\sc S.K. Agrawal, B.C. Fabien}, ``Optimization of Dynamical
Systems'', {\sl Solid Mechanics and Its Applications} Vol. {\bf
70}, Kluwer Academic Publishers, 1999.

\bibitem{BOV-2002}
{\sc G. Blankenstein, R. Ortega, A.J. Van der Schaft},
``The matching conditions of controlled Lagrangians and
IDA-passitivity based control",
 {\sl Internat. J. Control}{\bf 75}(9) (2000) 645-665.

\bibitem{B} {\sc A. M. Bloch:}  {\sl Nonholonomic mechanics and control}. Interdisciplinary Applied Mathematics, 24. Systems and
Control. Springer-Verlag, New York, 2003.

\bibitem{BLM-2000}
{\sc A. M. Bloch, N.E. Leonard, J.E. Marsden}, ``Controlled
Lagrangians and the Stabilization of Mechanical Systems I:
The First Matching Theorem"
 {\sl IEEE Trans. Aut. Cont.}{\bf 45}(12) (2000) 2253-2270.

\bibitem{Br}
{\sc T.J. Bridges, S. Reich}, ``Numerical methods for Hamiltonian PDEs'',
 {\sl J. Phys A. Math. Gen.}
{\bf 39} (2006) 5287--5320.

\bibitem{CLMM-2002}
{\sc J. Cort\'es, M. de Le\'on, D. Mart\'\i n de Diego, S.
Mart\'\i nez}, ``Geometric description of vakonomic and
nonholonomic dynamics. Comparison of solutions'', {\sl SIAM J.
Control and Optimization} {\bf 41}(5)  (2002) 1389-1412.

\bibitem{CMC-2002}
{\sc J. Cort\'es, S. Mart\'\i nez, F. Cantrijn}, ``Skinner-Rusk
approach to time-dependent mechanics'', {\sl Phys. Lett. A} {\bf
300} (2002) 250-258.

\bibitem{DI-2003}
{\sc M. Delgado-T\'ellez, A. Ibort}, ``A Panorama of Geometrical
Optimal Control Theory'', {\sl Extracta Mathematicae} {\bf
18}(2), 129--151 (2003).

\bibitem{ELMMR-04}
{\sc A. Echeverr\'\i a-Enr\'\i quez, C. L\'opez, J. Mar\'\i
n-Solano, M.C. Mu\~noz-Lecanda, N. Rom\'an-Roy},
``Lagrangian-Hamiltonian unified formalism for field theory'',
{\sl J. Math. Phys.} {\bf 45}(1) (2004) 360-385.

\bibitem{EMR-91}
{\sc A. Echeverr\'\i a-Enr\'\i quez, M.C. Mu\~noz-Lecanda, N.
Rom\'an-Roy}, ``Geometrical setting of time-dependent regular
systems. Alternative models'', {\sl Rev. Math. Phys.} {\bf 3}(3)
(1991) 301-330.

\bibitem{GM-05}
{\sc X. Gr\`acia, R. Mart\'\i n}, ``Geometric aspects of
time-dependent singular differential equations'', {\sl Int. J.
Geom. Methods Mod. Phys.} {\bf 2}(4) (2005) 597-618.

\bibitem{Ku-tdms}
{\sc R. Kuwabara}, ``Time-dependent mechanical symmetries and
extended Hamiltonian systems'', {\sl Rep. Math. Phys.} {\bf 19}
(1984) 27-38.

\bibitem{LMM-2002}
{\sc M. de Le\'on, J.C. Marrero, D. Mart\'\i n de Diego}, ``A new
geometrical setting for classical field theories'', {\sl
Classical and Quantum Integrability}. Banach Center Pub. {\bf 59},
Inst. of Math., Polish Acad. Sci., Warsawa (2002) 189-209.

\bibitem{LMM-2007}
{\sc M. de Le\'on, J.C. Marrero, D. Mart\'\i n de Diego}, ``Some
applications of semi-discrete variational integrators to classical
field theories'', to appear in {\sl Qualitative Theory and
Dynamical Systems}.

\bibitem{LR}
 {\sc M. de Le\'on, P.R. Rodrigues},
{\sl  Methods of Differential Geometry in Analytical Mechanics},
North-Holland Math. Ser. 152, Amsterdam, 1989.

\bibitem{MS-98}
{\sc L. Mangiarotti, G. Sardanashvily}, ``Gauge Mechanics'',
{\sl World Scientific}, Singapore, 1998.

\bibitem{muller}
{\sc P.C. M\"{u}ller}, ``Stability and optimal control of
nonlinear descriptor systems: A survey". {\sl Appl. Math.
Comput. Sci.} {\bf 8}(2) (1998) 269--286.

\bibitem{M-1999}
{\sc P.C. M\"{u}ller}, ``Linear-Quadratic Optimal Control
of descriptor systems". {\sl J. Braz. Soc. Mech. Sci.}
{\bf 21}(3) (1999) 423-432.

\bibitem{P62}
{\sc L. S. Pontryagin, V. G. Boltyanski, R. V. Gamkrelidze
and E. F. Mischenko}, \textit{The Mathematical Theory of Optimal
Processes}, Interscience Publishers, Inc., New York 1962.

\bibitem{Ra1}
{\sc M. F. Ra\~nada}, ``Extended Legendre transformation approach
to the time-dependent Hamiltonian formalism'', {\sl J. Phys. A:
Math. Gen.} {\bf 25} (1992) 4025-4035.

\bibitem{RRS}
{\sc A.M. Rey, N. Rom\'{a}n-Roy, M. Salgado},
``G\"{u}nther's formalism in classical
field theory: Skinner-Rusk approach and the evolution operator'',
{\sl J. Math. Phys.} {\bf 46}(5) (2005) 052901.

\bibitem{St-2005}
{\sc J. Struckmeier}, ``Hamiltonian dynamics on the symplectic
extended phase space for autonomous and non-autonomous systems'',
{\sl J. Phys. A: Math. Gen.} {\bf 38} (2005) 1275--1278.

\bibitem{Sa-89}
{\sc D.J. Saunders}, {\sl The Geometry of Jet Bundles}, London
Math. Soc. Lect. Notes Ser. {\bf 142}, Cambridge, Univ. Press,
1989.

\bibitem{SR-83}
{\sc R. Skinner, R. Rusk}, Generalized Hamiltonian dynamics I:
Formulation on $T^*Q\bigoplus TQ$'', {\sl J. Math. Phys.} {\bf 24}
(1983) 2589-2594.

\bibitem{Tulczy1}
{\sc W.M. Tulczyjew}, ``Hamiltonian systems, Lagrangian systems
and the Legendre transformation'', {\sl Symposia Mathematica}
{\bf 16} (1974) 247--258.
\end{small}

\end {thebibliography}
\end{document}